\documentclass[conference]{IEEEtran}
\usepackage{fancyhdr}
\usepackage{tikz}
\usepackage{amsmath}
\usepackage{amsfonts} 
\usepackage{mathrsfs} 
\usepackage{booktabs} 
\usepackage{enumitem}
\usepackage{graphicx}
\usepackage{fixltx2e}
\usepackage{url}
\usepackage{color}
\usepackage{listings}
\usepackage{xspace}
\usepackage{fancyvrb}
\usepackage{verbatim}
\usepackage[ruled,vlined,linesnumbered]{algorithm2e}
\usepackage{comment}
\usepackage{subfigure}
\usepackage{caption}
\usepackage{amssymb}
\usepackage{indentfirst}
\usepackage{setspace}
\usepackage[normalem]{ulem}
\usepackage{balance}

\newcommand{\cb}{\textcolor{blue}}
\definecolor{darkgreen}{rgb}{0,0.5,0}

\newcommand{\yjin}[1]{\cb{[yjin: #1]}}

\newcommand{\ppg}{PPG\xspace}

\newcommand{\nomg}{PSG\xspace}

\newcommand{\nodg}{PPG\xspace}
\newcommand{\nsv}{non-scalable vertex\xspace}
\newcommand{\nsvs}{non-scalable vertices\xspace}
\newcommand{\iv}{abnormal vertex\xspace}
\newcommand{\ivs}{abnormal vertices\xspace}
\newcommand{\para}[1]{\noindent \textbf{#1 }}

\newcommand{\ourtool}{\textsc{ScalAna}\xspace}
\newcommand{\graph}{program structure graph\xspace}

\newcommand{\dob}[1]{\textbf{#1}}

\newcommand{\hl}[2]{{{\texttt{\textcolor{#1}{#2}}}}}
\newcommand{\hlb}[2]{\hl{#1}{\dob{#2}}}

\definecolor{stcolor}   {rgb}{.0,.0,.0}
\definecolor{kwcolor}   {rgb}{.2,.2,.9}
\definecolor{typecolor} {rgb}{.2,.2,.9}
\definecolor{funccolor} {rgb}{.5,.2,.2}
\definecolor{varcolor}  {rgb}{.2,.2,.5}
\definecolor{cmtcolor}  {rgb}{.2,.5,.2}
\definecolor{constcolor}{rgb}{.7,.2,.7}
\definecolor{backcolor} {rgb}{.8,.8,.8}

\newcommand{\kw}[1]     {\textbf{\hl{kwcolor}{#1}}}

\newcommand{\func}[1]   {\textbf{\hlb{funccolor}{#1}}}
\newcommand{\var}[1]    {\textbf{\hl{stcolor}{#1}}}

\newcommand{\const}[1]  {\textbf{\hl{constcolor}{#1}}}

\def\BibTeX{{\rm B\kern-.05em{\sc i\kern-.025em b}\kern-.08em
    T\kern-.1667em\lower.7ex\hbox{E}\kern-.125emX}}

\usepackage{filecontents}

\begin{document}

\date{}


\title{\textsc{ScalAna}: Automating Scaling Loss Detection \\ with Graph Analysis}

\author{
{\rm Yuyang Jin}$^*$,
{\rm Haojie Wang}$^*$,
{\rm Teng Yu}$^*$,
{\rm Xiongchao Tang}$^*$,
{\rm Torsten Hoefler}$^\dag$,
{\rm Xu Liu}$^\ddag$,
{\rm Jidong Zhai}$^*$ \\
$^*$Tsinghua University, 
$^\dag$ETH Zürich, 
$^\ddag$North Carolina State University \\
\{jyy17, wang-hj18\}@mails.tsinghua.edu.cn,
yuteng@tsinghua.edu.cn,
tomxice@gmail.com,\\
htor@inf.ethz.ch,
xliu88@ncsu.edu,
zhaijidong@tsinghua.edu.cn
} 



\maketitle


\begin{abstract}

Scaling a parallel program to modern supercomputers is challenging due to inter-process communication, Amdahl's law, and resource contention. 
Performance analysis tools for finding such scaling bottlenecks either base on profiling or tracing. Profiling incurs low overheads but does not capture detailed dependencies needed for root-cause analysis. Tracing collects all information at prohibitive overheads.

In this work, we design \ourtool that uses static analysis techniques to achieve the best of both worlds - it enables the analyzability of traces at a cost similar to profiling. \ourtool first leverages static compiler techniques to build a {\em Program Structure Graph}, which records the main computation and communication patterns as well as the program's control structures. At runtime, we adopt lightweight techniques to collect performance data according to the graph structure and generate a {\em Program Performance Graph}. With this graph, we propose a novel approach, called \emph{backtracking root cause detection}, which can automatically and efficiently detect the root cause of scaling loss.
We evaluate \ourtool with real applications. Results show that our approach can effectively locate the root cause of scaling loss for real applications and incurs 1.73\% overhead on average for up to 2,048 processes. We achieve up to 11.11\% performance improvement by fixing the root causes detected by \ourtool on 2,048 processes.

\end{abstract}


\section{Introduction}
A decade after Dennard scaling ended and clock frequencies have stalled, increasing core count remains the only option to boost computing power. Top-ranked supercomputers~\cite{top500} already contain millions of processor cores, such as ORNL's Summit with 2,397,824 cores, LLNL's Sierra with 1,572,480 cores, and Sunway TaihuLight with 10,649,600 cores.
This unprecedented growth in the last years shifted the complexity to the developers of parallel programs, for which scalability is a main concern now.
Unfortunately, not all parallel programs have caught up with this trend and cannot efficiently use modern supercomputers, mostly due to their poor scalability~\cite{shi2012program,liu2015scaanalyzer}.

Scalability bottlenecks can have a multitude of reasons ranging from issues with locking, serialization, congestion, load imbalance, and many more~\cite{pearce2019exploring,schmidl2016openmp}. They often manifest themselves in synchronization operations and finding the exact root cause is hard. Yet, with the trend towards larger core count continuing, scalability analysis of parallel programs becomes one of the most important aspects of modern performance engineering. Our work squarely addresses this topic for large-scale parallel programs.

\begin{table}[h!]
\caption{Qualitative performance and storage analysis on state-of-the-art and \ourtool running NPB-CG with CLASS C for 128 processes~\cite{npb}}
\footnotesize
\begin{tabular}{p{1.7cm}p{1.8cm}p{1.9cm}p{1.6cm}}
\toprule  
Tools& Approaches &Time Overhead & Storage Cost\\
\midrule  
Scalasca~\cite{geimer2010scalasca} & Tracing-based & 25.3\% (w/o I/O)& 6.77 GB\\
HPCToolkit~\cite{adhianto2010hpctoolkit} & Profiling-based & 8.41\%& 11.45 MB \\
\ourtool & Graph-based & 3.53\%& 314 KB\\
\bottomrule 
\end{tabular}
\label{tab:intro-cost-compare}
\end{table}

Researchers have made great efforts in scalability bottleneck identification using three fundamental approaches: application profiling, tracing, and modeling.

\emph{Profiling-based approaches}~\cite{vetter2005mpip,tallent2010scalable,tallent2009diagnosing} collect statistical information at runtime with low overhead.
Summarizing the data statistically loses important information such as the order of events, control flow, and possible dependence and delay paths. Thus, such approaches can only provide a coarse insight into application bottlenecks and substantial human efforts are required to identify the root cause of scaling issues.

\emph{Tracing-based approaches}~\cite{itac,zhai2010phantom,geimer2010scalasca,linford2017performance} capture performance data as time series, which allows tracking dependence and delay sequences to identify root causes of scaling issues.
Their major drawback is the often prohibitive storage and runtime overhead of the detailed data logging. Thus, such tracing-based analysis can often not be used for large-scale programs. For example, we show the performance and storage overhead of the NPB-CG running with 128 processes comparing with tracing and profiling in Table~\ref{tab:intro-cost-compare} (Note that it is a single run for overhead comparison but not a typical use-case for scalability bottleneck identification.). 
Both profiling-based approaches and tracing-based approaches can use sampling techniques to reduce overhead but with a certain accuracy loss.


\emph{Modeling-based approaches}~\cite{yin2016discovering,yin2016joint,bhattacharyya2014pemogen,calotoiu2013using,wolf2016automatic,beckingsale2017apollo,linford2009multi} can also be used to identify scalability bottlenecks with low runtime overhead. 
However, building accurate performance models often requires significant human efforts and skills. Furthermore, establishing full performance models for a complex application with many input parameters requires many runs and prohibitively expensive~\cite{7776507}.
Thus, we conclude that identifying scalability bottlenecks for large-scale parallel programs remains an important open problem.

\begin{figure*}[ht]
    \centering
    \includegraphics[width=\linewidth]{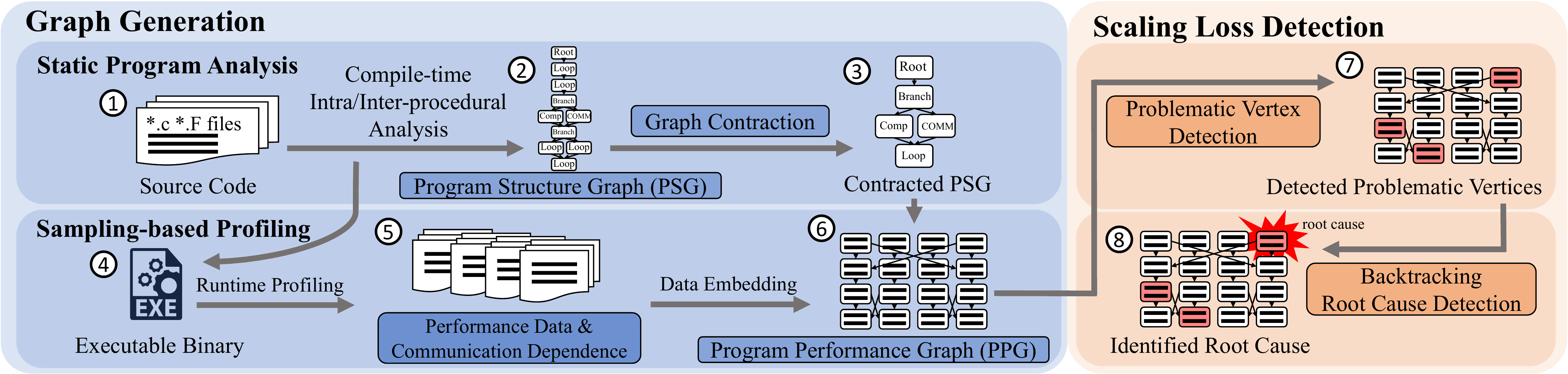}
    \caption{Overview of \ourtool}
    \label{fig:overview}
    \vspace{-1.5em}
\end{figure*}

To accurately identify scalability problems with low effort and overhead, we consider the program structure during data profiling. \ourtool combines \textbf{static program analysis with dynamic sampling-based profiling} into a light-weight mechanism to automatically identify the root cause of scalability problems for large-scale parallel programs.
We utilize an intra- and inter-procedural analysis of the source-code structure and record dynamic message matching at runtime to establish an efficient dependence graph of the overall execution. In particular, \ourtool is able to detect latent scaling issues in complex parallel programs where the delay will propagate to the other processes after several time steps through massive communication dependence. In summary, there are three main contributions in our work:


\begin{itemize}[leftmargin = 10pt]
    \item We design a fine-grained {\em Program Structure Graph} (PSG) that represents a compressed form of all program dependence within and across parallel processes. Then we generate \emph{Program Performance Graph} (PPG) by enhancing \nomg for each execution combining static compile-time analysis with light-weight runtime profiling.
    
    \item Based on the \ppg, we design a location-aware algorithm to detect problematic vertices with scaling issues. Combining inter-process dependence chains, we further propose a novel graph analysis algorithm, called \emph{backtracking root cause detection}, to find their root cause in source code. 
    
    \item We implement a light-weight performance tool named \ourtool\footnote{\ourtool is available at: \textit{\url{https://github.com/thu-pacman/SCALANA}}.}, and evaluate it with real applications. Results show that \ourtool can effectively and automatically identify the root cause of scalability problems.
\end{itemize}

We evaluate \ourtool with both benchmarks and real applications. Experimental results show that our approach can identify the root cause of scalability problems for real applications more accurately and effectively comparing with HPCToolkit~\cite{adhianto2010hpctoolkit} and Scalasca~\cite{geimer2010scalasca}.
\ourtool only incurs 1.73\% overhead on average for evaluated programs up to 2,048 processes. We achieve up to 11.11\% performance improvement by fixing the root causes detected by \ourtool on 2,048 processes.



\section{Design Overview}
\label{sec:overview}
One main innovation of \ourtool is to build a {\em Program Structure Graph} (PSG) at compile time and use it during runtime to minimize tracing overheads. The PSG captures the main computation and communication patterns that can be extracted statically from a parallel program. During the execution, \ourtool collects light-weight performance data as PSG vertex attributes as well as communication dependence between different processes and finally forms a {\em Program Performance Graph} (PPG). Another innovation of \ourtool is that we leverage the features of the generated \ppg to locate problematic vertices and then we use graph analysis to automatically identify the root cause of scaling issues in the source code.

In general, \ourtool consists of two main modules, \textit{graph generation} and \textit{scaling loss detection}. Figure~\ref{fig:overview} shows the high-level workflow of our system. Graph generation contains two phases, static program analysis and sampling-based profiling. Static program analysis is done at compile time while the sampling-based profiling is performed at runtime. We use the LLVM compiler~\cite{llvm} to automatically build a \nomg. Each vertex on the \graph is corresponding to a code snippet in the source code. The scaling loss detection is an offline module, which includes problematic vertex detection and root-cause analysis. We describe several key steps of these two modules below.

\textbf{Graph Generation}
	\begin{itemize}[leftmargin = 10pt]
		\item \textit{Program Structure Graph (PSG).} The input of this module is the source code of a parallel program. Through an intra- and inter-procedural static analysis of the program, we get a preliminary \textit{Program Structure Graph} (Section~\ref{subsub:psg}). 
		\item \textit{Graph Contraction.} In this step, we remove unnecessary edges in the PSG and merge several small vertices into a large vertex to reduce scalability analysis overhead (Section~\ref{subsub:pruning}). 
		\item \textit{Performance Data and Communication Dependence.} To effectively detect the scalability bottleneck, we leverage sampling techniques to collect the performance data for each vertex of the PSG and communication dependence data with different numbers of processes (Section~\ref{subsec:runtime-graph-supplement}). 
		\item \textit{Program Performance Graph (PPG).} To analyze the interplay of computation and communication among different processes, we further generate a \textit{Program Performance Graph} based on per-process PSGs (Section~\ref{subsec:extract-dependence-graph}). 
	\end{itemize}

\textbf{Scaling Loss Detection}
	\begin{itemize}[leftmargin = 10pt]
		\item \textit{Problematic Vertex Detection.} According to the structure of the acquired \ppg, we design a location-aware detection approach to identify all problematic vertices (Section~\ref{sub:location-aware-detect}).
		\item \textit{Backtracking Root Cause Detection.} 
		Combined with identified problematic vertices, we propose a backtracking algorithm on top of the \ppg and identify all the paths covering problematic vertices, which can help locate the root cause of the scaling issues (Section~\ref{sub:back-propa-root}).
	\end{itemize}



\begin{figure}[h]
	\centering
	\includegraphics[width=\linewidth]{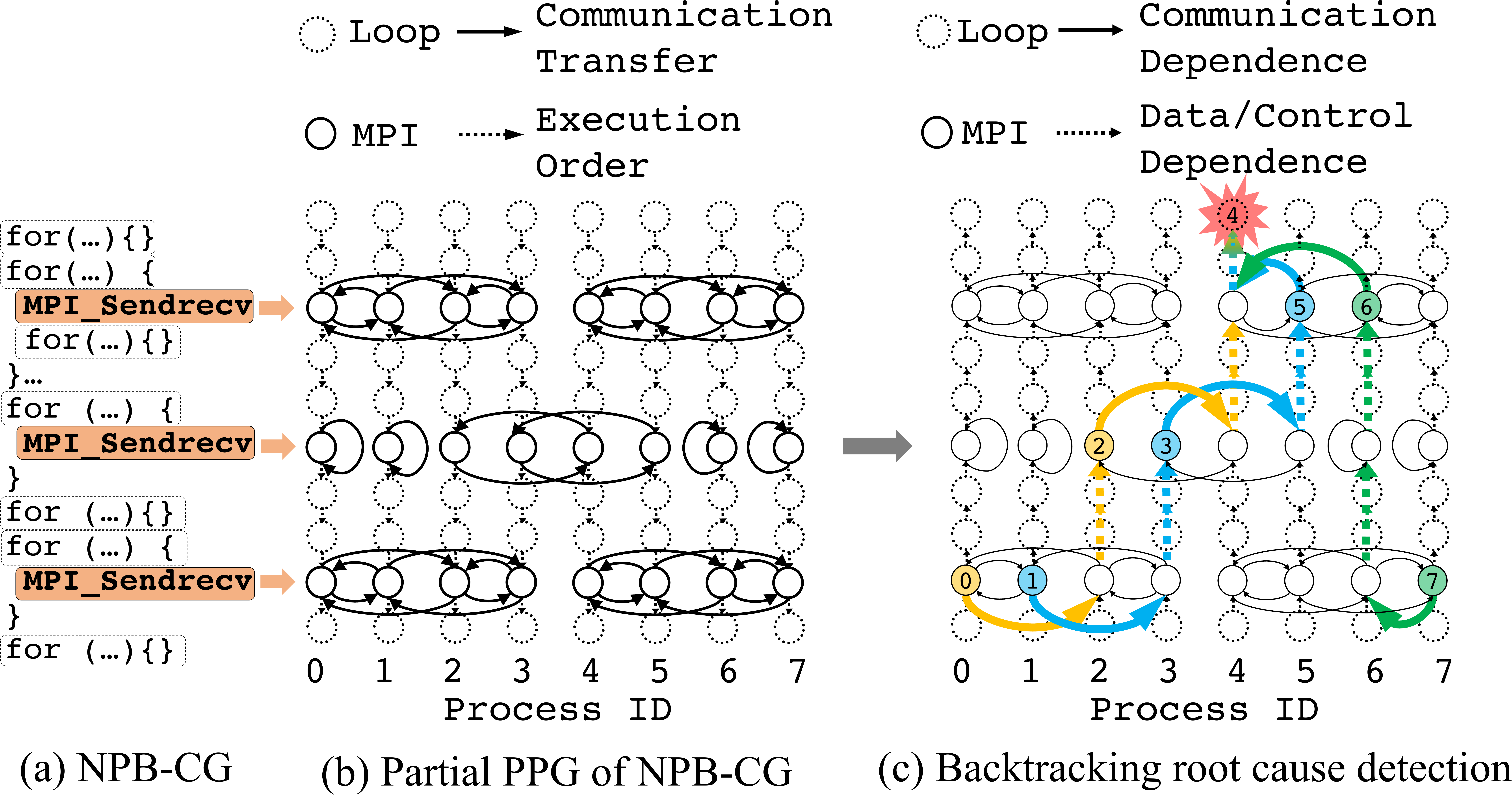}
	\caption{Motivating example with NPB-CG}
	\label{fig:cg-comm-pattern}
\end{figure}

We give an example to show how our approach is used to detect scaling loss. Figure~\ref{fig:cg-comm-pattern} shows the code snippet of NPB-CG~\cite{npb} and the partial \ppg of NPB-CG generated by \ourtool (due to space limitation, we only draw the \ppg for 8 processes). We manually inject a delay in one process (process $4$), which causes a scaling loss on the Tianhe-2 system (49.4 seconds at 1,024 processes vs. 49.5 seconds at 2,048 processes). Tracing-based approaches like Scalasca~\cite{geimer2010scalasca} and Vampir~\cite{nagel1996vampir} generate more than 250 GB of trace data. 
Due to the covert performance issue mixed by data, control, and inter-process communication dependence, we observe that the traditional profiling-based tool, like HPCToolkit~\cite{adhianto2010hpctoolkit}, needs significant human efforts to identify the accurate root cause for this case.

In \ourtool, we leverage both the static and dynamic analysis to build a holistic \ppg that records program execution order and data flow as well as inter-process communication transfer, as shown in Figure~\ref{fig:cg-comm-pattern}(b).
With our detection algorithm, we first identify some problematic vertices in this graph. For example, the vertices are marked with red, blue, yellow, or green color. In general, a problematic vertex is a vertex with unusual performance relative to other vertices.
Then we perform a backtracking root cause detection on the \ppg of NPB-CG, as shown in Figure~\ref{fig:cg-comm-pattern}(c). 
Through backward traversing this graph, we can detect the red vertex of process $4$ is the root cause through a path of vertices that traverses different processes. 

In summary, \ourtool is a programmer-oriented scalability analysis tool, which takes input as the source code of a parallel program, detects the root cause of scaling bottlenecks and reports back to the programmer which lines of the source code cause the problems to guide further optimization on the program.

\section{Graph Generation}
\label{sec:graph}

In this section, we describe how we automatically build an appropriate representation to reflect the main computation and communication characteristics for a given parallel program in detail. 
Our approach mainly relies on a static program analysis module. It also incorporates a sampling-based profiling module to handle input-dependent information.

\subsection{Static Program Structure Graph Construction}
\label{subsub:psg}

In general, the static analysis module is in charge of building a per-process \nomg, which can be regarded as a sketch of a parallel program. In a PSG, the vertices represent main computation and communication components as well as program control flow. 
The edges represent their execution order based on both data and control flow.
We group the vertices into different types, including \textit{Branch}, \textit{Loop}, \textit{Function call}, and \textit{Comp}, among which, \textit{Comp} is a collection of computation instructions while the others are basic program structures.

There are three main phases to build a \nomg statically: intra-procedural, inter-procedural analysis, and graph contraction. During the intra-procedural analysis, we firstly build a local \nomg for each function. And then through an inter-procedural algorithm, we acquire a complete \nomg, which will be further refined by graph contraction.


\para{Intra-procedural Analysis} During the intra-procedural analysis phase, we build a local \nomg for each procedure. The basic idea is that we traverse the control flow graph of the procedure at the level of the intermediate representation (IR) of the program, identify loops, branches, and function calls, and then connect these components based on their dependence to form a per-function local graph.

\para{Inter-procedural Analysis} Inter-procedural analysis is to combine all the local PSGs into a complete graph. We start by analyzing the program's call graph (PCG), which contains all calling relationships between different functions. And then we perform a top-down traversal of the PCG from the \texttt{main} function and replace all user-defined functions with their local PSGs. For MPI function calls, we just keep them. For indirect function calls, we need to process them after collecting certain function call relationships at runtime. 
For recursive function calls, their edges are similar to the recursive call edges in the PCG, which means that a circle is formed in the PSG. After the static analysis, the runtime performance data will be attached to these vertices with extra call-stack information for further analysis.

\para{PSG Contraction}
\label{subsub:pruning}
The PSGs generated in the above step are normally too large to be applied efficiently for real applications since we need to create corresponding vertices for any loop and branch in their source code. However, the workload of some vertices can be ignored as collecting performance data for these vertices only introduces large overhead without benefits. To address this problem, \ourtool performs graph contraction to reduce the size of the generated PSG.

The rules of contraction affect the granularity of the graph and the representation of communication and computation characteristics. 
Considering that communication is normally the main scalability bottleneck for parallel programs, \ourtool preserves all MPI invocations and related control structures.
For computation vertices in the PSG, we merge continuous vertices into a larger vertex. 
Specifically, for the structures that do not include MPI invocations, we only preserve \textit{Loop} because computation produced by loop iterations may dominate performance. In addition, \ourtool allows a user-defined parameter, \texttt{MaxLoopDepth}, as a threshold to limit the depth of nested loops and keep the graph condensed.


\begin{figure}[h!]
\centering
\footnotesize
\begin{Verbatim}[numbers=left,numbersep=2mm,xleftmargin=3mm,commandchars=\\\{\}]
\kw{int} \func{main}()\{
    \kw{for} (\kw{int} \var{i} = \const{0}; \var{i} < \var{N}; ++\var{i})      //Loop 1
        \var{A}[\var{i}] = \func{rand}();
        \kw{for} (\kw{int} \var{j} = \const{0}; \var{j} < \var{i}; ++\var{j})  //Loop 1.1
            \var{sum} += \var{A}[\var{j}];
        \kw{for} (\kw{int} \var{k} = \const{0}; \var{k} < \var{i}; ++\var{k})  //Loop 1.2
            \var{product} *= \var{A}[\var{k}];
    \func{foo}(); 
    \func{MPI_Bcast}(...);
\}
\kw{void} \func{foo}() \{
    \kw{if} (\var{myRank} \% \const{2} == \const{0}) 
        \func{MPI_Send}(...); 
    \kw{else}
        \func{MPI_Recv}(...); 
\}
\end{Verbatim}
\setlength{\abovecaptionskip}{0mm}
\caption{An MPI program example} 
\label{fig:example}
\end{figure}

\begin{figure}[h!]
	\centering
	\includegraphics[width=\linewidth]{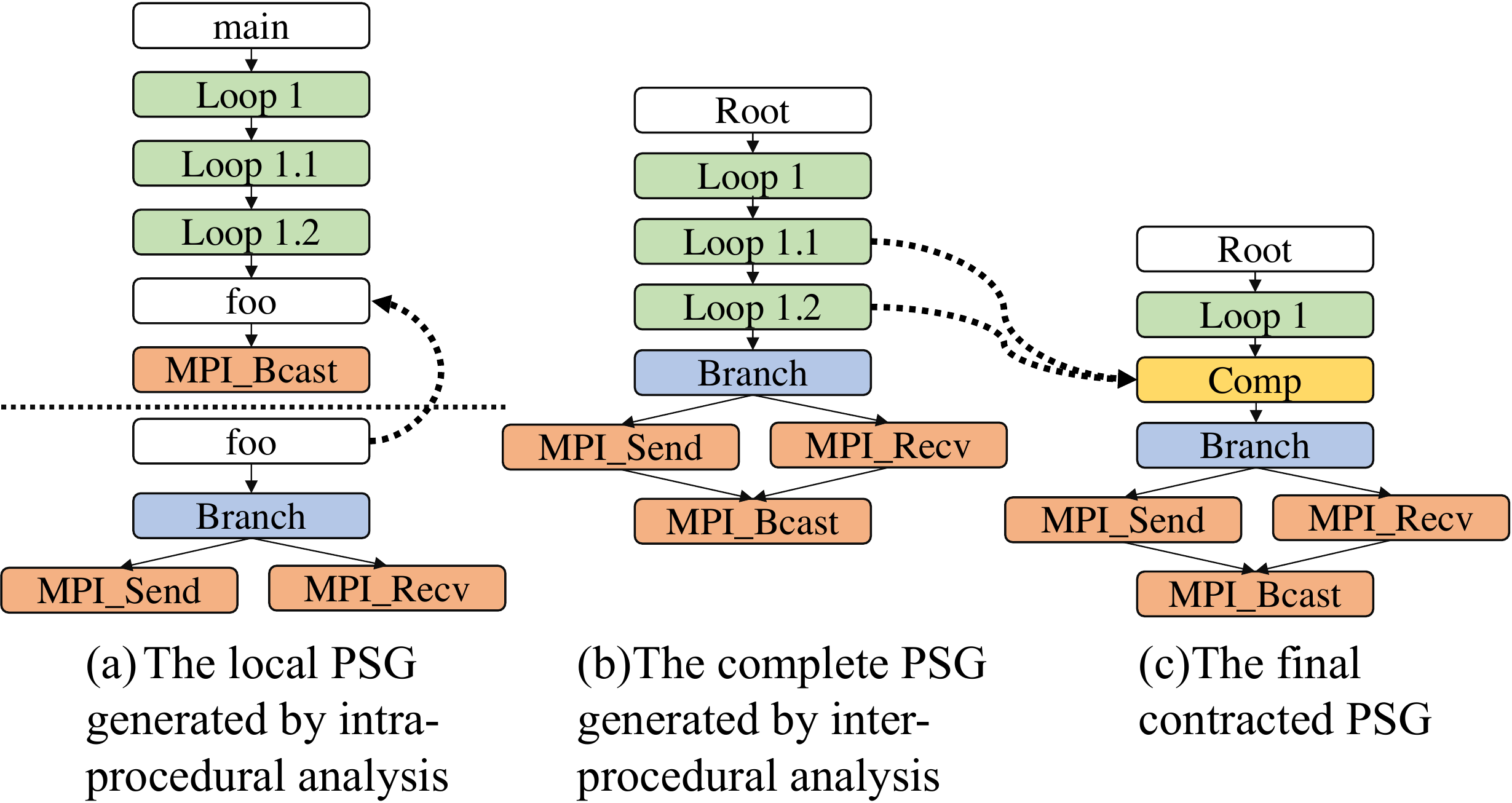}
	\caption{Static Program Structure Graph Generation}
	\label{fig:psg}
\end{figure}

For instance, Figure~\ref{fig:example} shows a simple MPI program example with two functions. Figure~\ref{fig:psg}(a) shows its local PSGs for each function after the intra-procedural analysis. Figure~\ref{fig:psg}(b) shows a complete \nomg after the inter-procedural analysis. Figure~\ref{fig:psg}(c) shows the contracted PSG after merging the sequential \texttt{Loop1.1} and \texttt{Loop1.2} when \texttt{MaxLoopDepth} is set to 1.

\subsection{Sampling-Based Profiling}
\label{subsec:runtime-graph-supplement}
\ourtool is a hybrid approach. We design a sampling-based profiling module to annotate the PSG with profiling data and also refine it based on runtime information. The sampling-based profiling module includes performance profiling, inter-process dependence profiling, and indirect call analysis. Performance profiling is to collect and fill runtime metrics into the vertices of the graph to handle input-dependent information. We use a sampling technique for performance profiling (Section~\ref{subsec:perf-data-collection}). Inter-process dependence profiling is to connect per-function PSG into a larger graph (\ppg) that cannot be derived statically (Section~\ref{subsub:comm-dependence}).

\subsubsection{Associate Vertices with Performance Data}
\label{subsec:perf-data-collection}
We collect the performance data for each vertex of the PSG at runtime, which is essential for further analysis of scaling issues. Unlike traditional coarse-grained profiling approaches, \ourtool collects performance data according to the granularity of each PSG vertex. One main advantage is that we can combine the graph structure and performance data for more accurate performance analysis. Specifically, we associate each PSG vertex with a performance vector that records the execution time and key hardware performance data, such as cache miss rate and branch miss count. 

We use sampling techniques for performance profiling to collect metrics with very low overhead. 
We use PAPI~\cite{papi} for sampling and hardware performance data collection, which interrupts the program at regular clock cycles and records program call stack and related performance data. According to the program call stack information, we can associate performance data with the corresponding PSG vertex at the interruption point. 

\subsubsection{Graph-Guided Communication Dependence}
\label{subsub:comm-dependence}
During the static analysis, we derive data and control dependence within each process. At runtime, we need to further collect communication dependence between different processes for inter-process dependence analysis. 
Traditional tracing-based approaches record each communication operation and analyze their dependence, which causes large collection overhead and also huge storage cost~\cite{wu2011scalaextrap,noeth2009scalatrace}. We propose two key techniques to address this problem: sampling-based instrumentation and graph-guided communication compression. 

\para{Sampling-Based Instrumentation}
Full instrumentation always introduces large overheads. 
The dynamic program behavior may be missed if the instrumentation is recorded only once.
To reduce the runtime overhead and still capture the dynamic program behavior along with the program execution, we adopt a random sampling-based instrumentation technique~\cite{vetter2002dynamic}. 
A random number is generated every time when the instrumentation is executed. When the random number falls into an interval of the pre-defined threshold we record communication parameters.
The random sampling technique used here can avoid missing regular communication patterns as much as possible even if they change at runtime.

\para{Graph-Guided Communication Compression} 
A typical parallel program contains a large number of communication operations. Due to the redundancy between different loop iterations, we do not need to record all the communication operations. As the \nomg already represents the program's high-level communication structure, we can leverage this graph to reduce communication records. We only record communication operation parameters once for repeated communications with the same parameters of the recorded data, which can reduce the storage cost and ease the analysis of inter-process dependence.

We use PMPI~\cite{Mohr2011} in this work for effective communication collection, which does not need to modify the source code. For different communication types, we adopt different methods to collect their dependence. We distinguish three common classes of communication: (1) For collective communication, we should know which processes are involved in this communication. In MPI programs, we can use \texttt{MPI\_Comm\_get\_info} to acquire this information. (2) For blocking point to point communication, we should record the \textit{source} or \textit{dest} process and \textit{tag} directly. (3) For non-blocking communication, some information will not be available until final checking functions are invoked (such as \texttt{MPI\_Wait}). 

\begin{figure}[t]
	\centering
    \scriptsize
    \begin{Verbatim}[numbers=left,numbersep=2mm,xleftmargin=3mm,commandchars=\\\{\}]
\kw{map} <\kw{MPI_Request}*, \kw{pair}<\kw{int},\kw{int}>> \var{requestConverter};
\kw{int} \func{MPI_Irecv}(..., \kw{int} \var{source}, \kw{int} \var{tag},
                        ..., \kw{MPI_Request} *\var{request})\{
    \var{requestConverter}[\var{request}] = <\var{source},\var{tag}>;
    \kw{return} \func{PMPI_Irecv}(...);
\}
\kw{int} \func{MPI_Wait}(\kw{MPI_Request} *\var{request}, \kw{MPI_Status} *\var{status})\{
    \var{retval} = \func{PMPI_Wait}(\var{request}, \var{status});
    <\var{source},\var{tag}> = \var{requestConverter}[\var{request}];
    \kw{if} (\var{source} or \var{tag} is uncertain) \{
        \var{commSet}.\func{insert}(<\var{status.MPI_SOURCE},\var{status.MPI_TAG}>);
    \} \kw{else} \{
        \var{commSet}.\func{insert}(<\var{source}, \var{tag}>);
    \}
    \kw{return} \var{retval};
\}
    \end{Verbatim}
    \setlength{\abovecaptionskip}{0mm}
	\caption{Acquiring communication dependence for non-blocking communications}
	\label{fig:comm-set-irecv}
\end{figure}

We take \texttt{MPI\_Wait} after \texttt{MPI\_Irecv} as an example as shown in Figure~\ref{fig:comm-set-irecv}. Firstly, we store the \textit{source} process and \textit{tag} from the parameters associated to the \textit{request} in \texttt{MPI\_Irecv}. Then in \texttt{MPI\_Wait}, the \textit{source} and \textit{tag} corresponding to the \textit{request} are recorded into a communication dependence set. If the \textit{source} or \textit{tag} is uncertain, we acquire them from the parameter of \textit{status} in \texttt{MPI\_Wait}.

\subsubsection{Indirect Function Calls} 
Sometimes, the program call graph cannot be fully obtained by the static analysis due to indirect calls, such as function pointers. We need to collect the calling information of indirect calls at runtime and fill such information into the graph. We do necessary instrumentation before the entry and exit of indirect calls and link this information with real function calls with unique function IDs and then refine the PSG obtained after the inter-procedural analysis. 

\subsection{Program Performance Graph}
\label{subsec:extract-dependence-graph}
\begin{figure}[h!]
	\centering
	\includegraphics[width=\linewidth]{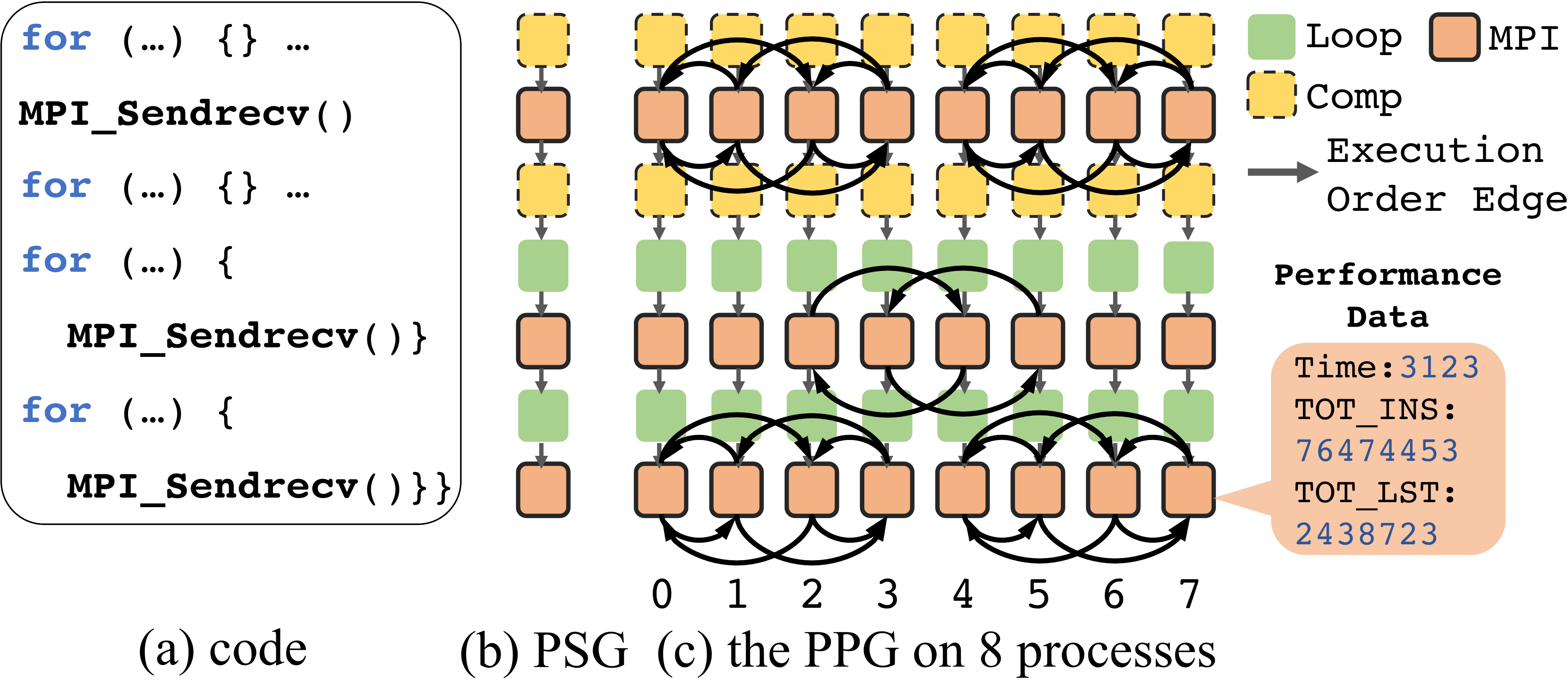}
	\caption{A PPG running with 8 processes}
	\label{fig:cg-inter-wait}
\end{figure}

After both the static program analysis and sampling-based profiling, we build a final {\em Program Performance Graph} (PPG). As each process shares the same source code, we can duplicate the PSG for all processes. Then we add inter-process edges based on communication dependence collected at the runtime analysis. For point to point communications, we match the sending and receiving processes. For collective communications, we associate all involved processes. Figure~\ref{fig:cg-inter-wait} shows a simplified final \ppg for an example program running with 8 processes.

Note that the final \ppg not only includes the data and control dependence for each process but also records the inter-process communication dependence. In addition, we also attribute key performance data for each vertex, which will be used for further scaling issue detection. For a given vertex in this graph, its performance can be affected by either its own computation patterns or the performance of other vertices connected through data and control dependence within one process as well as communication dependence between different processes. We describe how we locate the performance issue below. 
\section{Scaling Loss Detection}
\label{sec:analysis}

In this section, we describe how we leverage the acquired \ppg for effective and automatic scaling loss detection. Our approach consists of two key steps, location-aware problematic vertex detection and backtracking root cause identification. The former is to detect problematic vertices with poor scalability or abnormal behavior. The latter is to pinpoint the root cause of scaling loss problems. 

\subsection{Location-Aware Problematic Vertex Detection}
\label{sub:location-aware-detect}
One main advantage of our approach is that we have generated a final \ppg from a given program. Although the inter-process communication dependence may change with the different numbers of processes, the per-process PSG does not change with the problem size or job scale. Based on this observation, we propose a location-aware detection approach to identify problematic vertices. The core idea of our approach is that we compare the performance data of vertices in the \ppg which corresponds to the same vertex in the \nomg among different job scales (\nsv detection) and different processes for a given job scale (\iv detection). 

\begin{figure}[h]
	\centering
	\begin{minipage}[t]{\linewidth}
	    \centering
		\subfigure[A \nsv example]{
			\label{fig:scal-iss-ver}
			\includegraphics[width=0.46\linewidth]{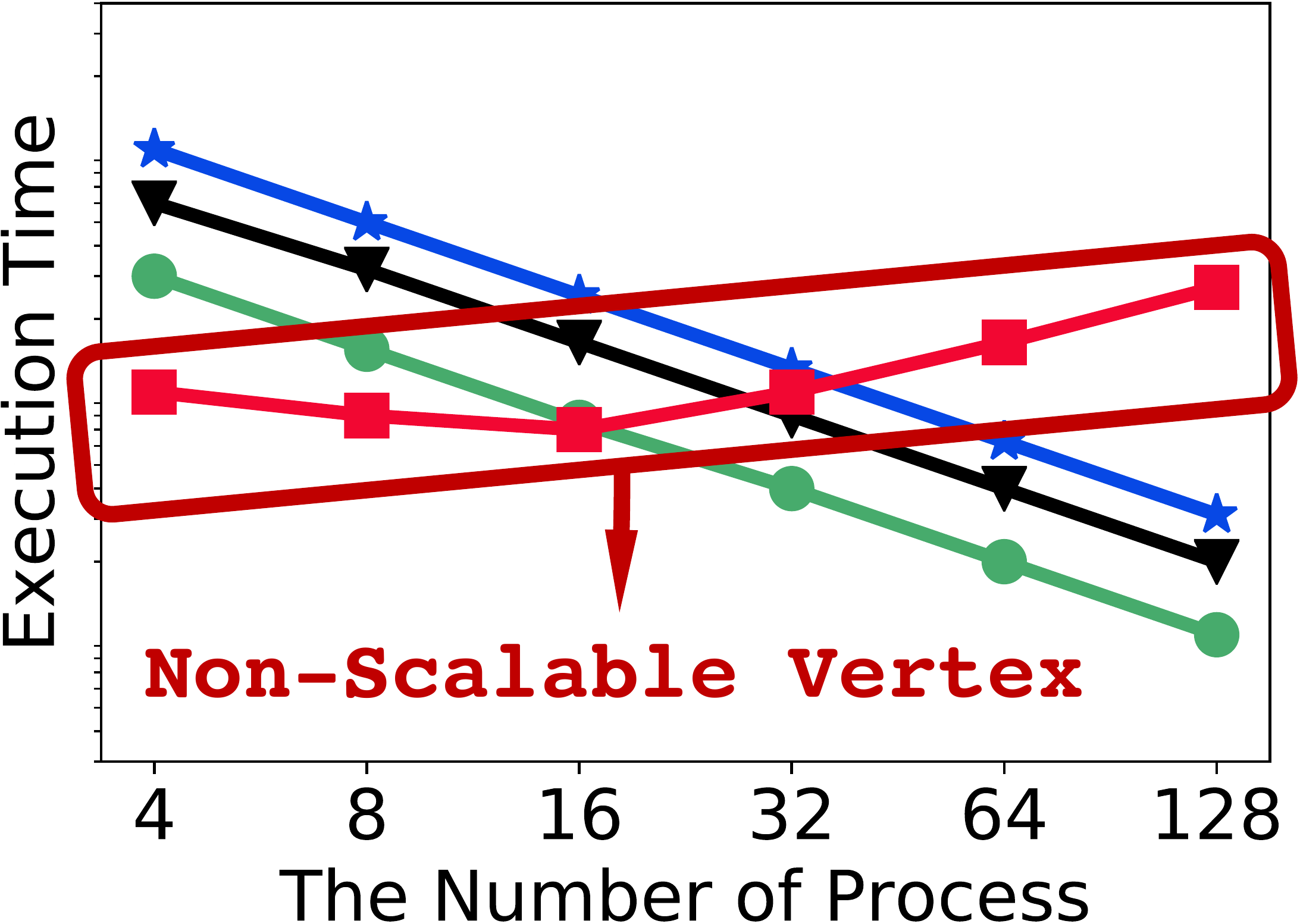} 
		}
		\subfigure[An \iv example]{
			\label{fig:abnor-ver}
			\includegraphics[width=0.46\linewidth]{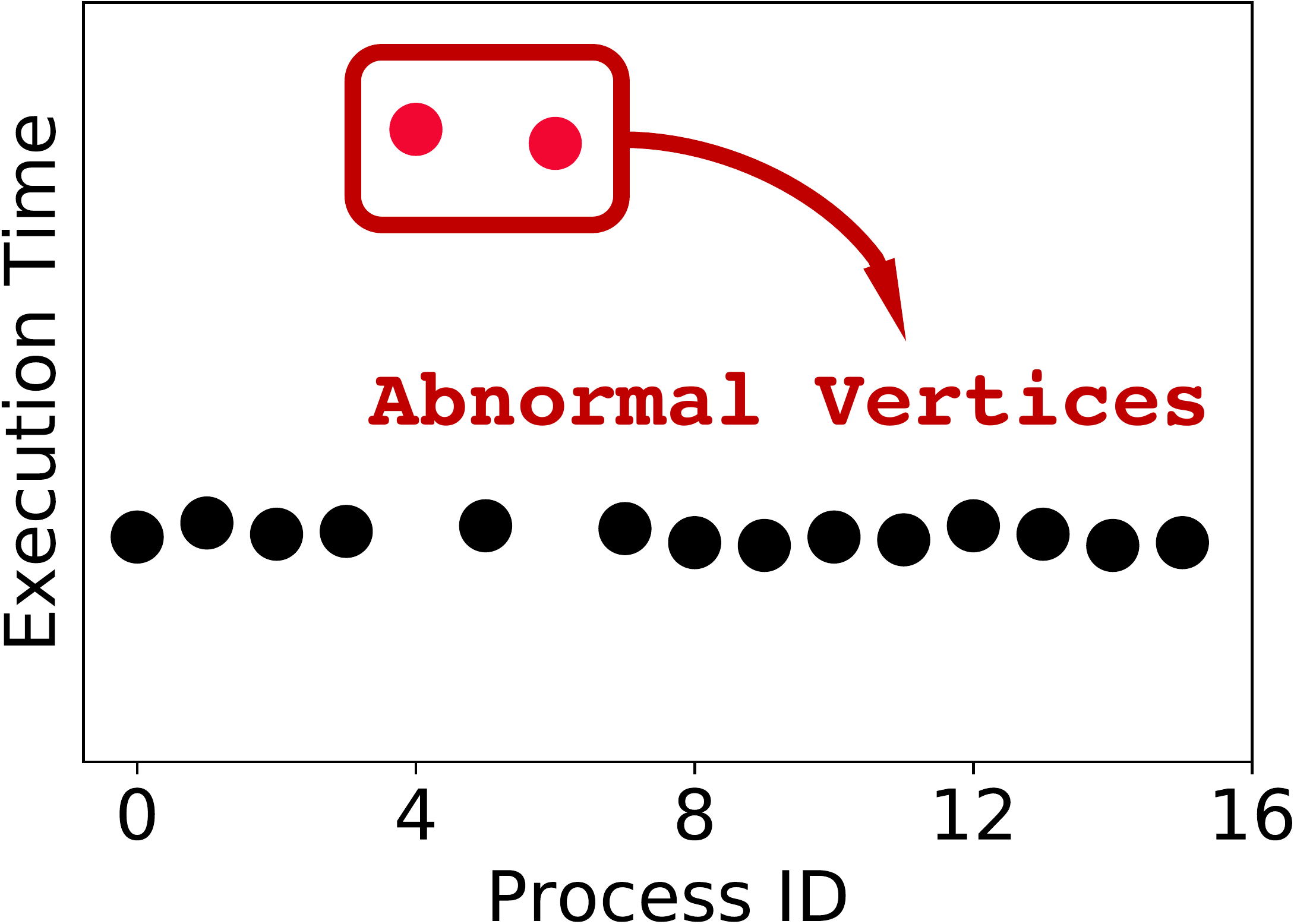} 
		}
	\end{minipage}
	\vspace{-0.5em}
	\caption{Two kinds of problematic vertices examples}
	\label{fig:defective-exa}
\end{figure}

\noindent \textbf{Non-Scalable Vertex Detection} The core idea is to find vertices in the \ppg whose performance (execution time or hardware performance data) shows an unusual slope comparing with other vertices when the number of processes increases. For instance, Figure~\ref{fig:scal-iss-ver} shows the change of the execution time of different vertices in a \nomg as the process count increases. The execution time of the vertex in the red line does not decrease like other vertices. When the execution time of these vertices accounts for a large proportion of the total time, they will become a scaling issue.

A challenge for \nsv detection is how to merge performance data from a large number of processes. The simplest strategy is to use the performance data for a particular process for comparison but this strategy may lose some information about other processes. Another strategy is to use the mean or median value of performance data from all processes and the performance variance among different processes to reflect load distribution. 
We can also partition all processes into different groups by clustering algorithms and then aggregate for each group. 
In our implementation, we test all strategies mentioned above and fit the merged data of different process counts with a log-log model~\cite{barnes2008regression}. With these fitting results, we sort all vertices by the changing rate of each vertex when the scale increases and filter the top-ranked vertices as the potential \nsvs.

\noindent  \textbf{Abnormal Vertex Detection} For a given job scale, we can also compare the performance data of the same vertex among different processes. Since for typical SPMD (Single Program Multi-Data) programs, the same vertex tends to execute the same workload among different processes. If a vertex has significantly different execution time, we can mark this vertex as a potential \iv. A lot of reasons can cause \ivs, even if we do not consider the effect of performance variance~\cite{tang2018vs}. For instance, a load balance problem can cause \ivs in some processes. We can also identify some communication vertices that have much larger synchronization overhead than other processes. Figure~\ref{fig:abnor-ver} shows the execution time of the vertices of all processes in a \ppg which correspond to the same vertex in the \nomg on 16 processes. Among them, process 4 and 6 take longer to execute than the others and yield the \ivs. 
\ourtool allows a user-defined threshold \texttt{AbnormThd} to distinguish both abnormal and normal vertices among different parallel processes. We discuss details in Section~\ref{sub:eval-cases}.

As shown in Figure~\ref{fig:cg-pagerank-result}, after the analysis of the above two steps, we mark some problematic vertices in the \ppg (vertices with blue and red color) generated in Figure~\ref{fig:cg-inter-wait}.

\subsection{Backtracking Root Cause Detection}
\label{sub:back-propa-root}
Furthermore, we need to connect the identified problematic vertices and find the causal relationship between them to locate the root cause of the scaling problem. In this work, we use graph analysis to propose a novel approach, named backtracking root cause detection to automatically report the line number of source code corresponding to the root cause. 

To do the backward traversal, first we need to reverse all edges to dependence edges.
The pseudo-code of \textit{backtracking root cause algorithm} is shown in Algorithm~\ref{alg:roocau}. Our algorithm starts from the \nsvs detected in the above step, then tracks backward through data/control dependence edges within a process and communication dependence edges among different processes until the root vertices or collective communication vertices are found. 
If unscanned \textit{Loop} or \textit{Branch} vertices are found during the backtrack, our algorithm will traverse only the control dependence edges but not the data dependence edges. For example, when a \textit{Loop} vertex is found, the traversal continues from the end vertex of this loop.
One observation is that a complex parallel program always contains a large number of dependence edges. So the search cost will be very high if we would not optimize. However, we do not need to traverse all the possible paths to identify the root cause. 
In \ourtool, we only preserve the communication dependence edge if a waiting event exists while we prune other communication dependence edges. The advantage of our approach is that we can reduce both searching space and false positives.
Finally, we get several causal paths that connect a set of \ivs. Further analysis of these identified paths will help application developers to locate the root cause. 

\begin{algorithm}[h!]
	{\fontsize{8pt}{9pt}\selectfont
	\caption{Backtracking Root Cause Algorithm}	
	\label{alg:roocau}
		\KwIn{A Program Performance Graph $\mathit{PPG}$, A Set of Non-Scalable Vertices $\mathbb{N}$, A Set of Abnormal Vertices $\mathbb{A}$. }
		\KwOut{A Set of Root Cause Paths $\mathbb{S}$. }
		\SetKwFunction{Begin}{Backtracking}
		\SetKwFunction{FMain}{Main}
		\SetKwProg{Fn}{Function}{:}{}
		\Fn{\FMain{}}{
		$\mathbb{S} \gets \varnothing$ \;
		$\mathbb{V} \gets \varnothing$ \tcp*[l]{Set of scanned vertices} 
		\ForAll {$\mathit{n} \in \mathbb{N}$} {
		    $\mathbb{P} \gets \varnothing$ \tcp*[l]{Root cause path} 
		    $\texttt{Backtracking}$ ( $\mathit{n}$, $\mathbb{P}$ ) \;
            Insert $\mathbb{P}$ into $\mathbb{S}$ \;
            Insert all $\mathit{v} \in \mathbb{P}$ into $\mathbb{V}$ \;
		}
		\ForAll (\tcp*[h]{Traverse the vertices that have not been scanned}) {$\mathit{a} \in \mathbb{A}$ and $\mathit{a} \notin \mathbb{V}$} {
		    $\mathbb{P} \gets \varnothing$\; 
		    $\texttt{Backtracking}$ ( $\mathit{a}$, $\mathbb{P}$ ) \;
            Insert $\mathbb{P}$ into $\mathbb{S}$ \;
		}
		\textbf{return} $\mathbb{S}$ \;
		}
		\Fn{\Begin{$\mathit{v}$, $\mathbb{P }$  }}{
		    \While {$\mathit{v}$ is not root or collective communication vertex} {
		        Insert $\mathit{v}$ into $\mathbb{P}$ \;
		        \uIf {$\mathit{v}$ is an \texttt{MPI} vertex}{
		            $\mathit{v} \gets$ the $\mathit{dest}$ vertex of inter-process communication dependence edge of $\mathit{v}$ \;
		        }\uElseIf{$\mathit{v}$ is an unscanned \texttt{LOOP} or \texttt{BRANCH} vertex}{
		            $\mathit{v} \gets$ the $\mathit{dest}$ vertex of control dependence edge of $\mathit{v}$\;
		        }\uElse{
		            $\mathit{v} \gets$ the $\mathit{dest}$ vertex of data dependence edge of $\mathit{v}$ \;
		        }
		    }
		}
	}

\end{algorithm}



\begin{figure}[h!]
	\centering
	\includegraphics[width=0.7\linewidth]{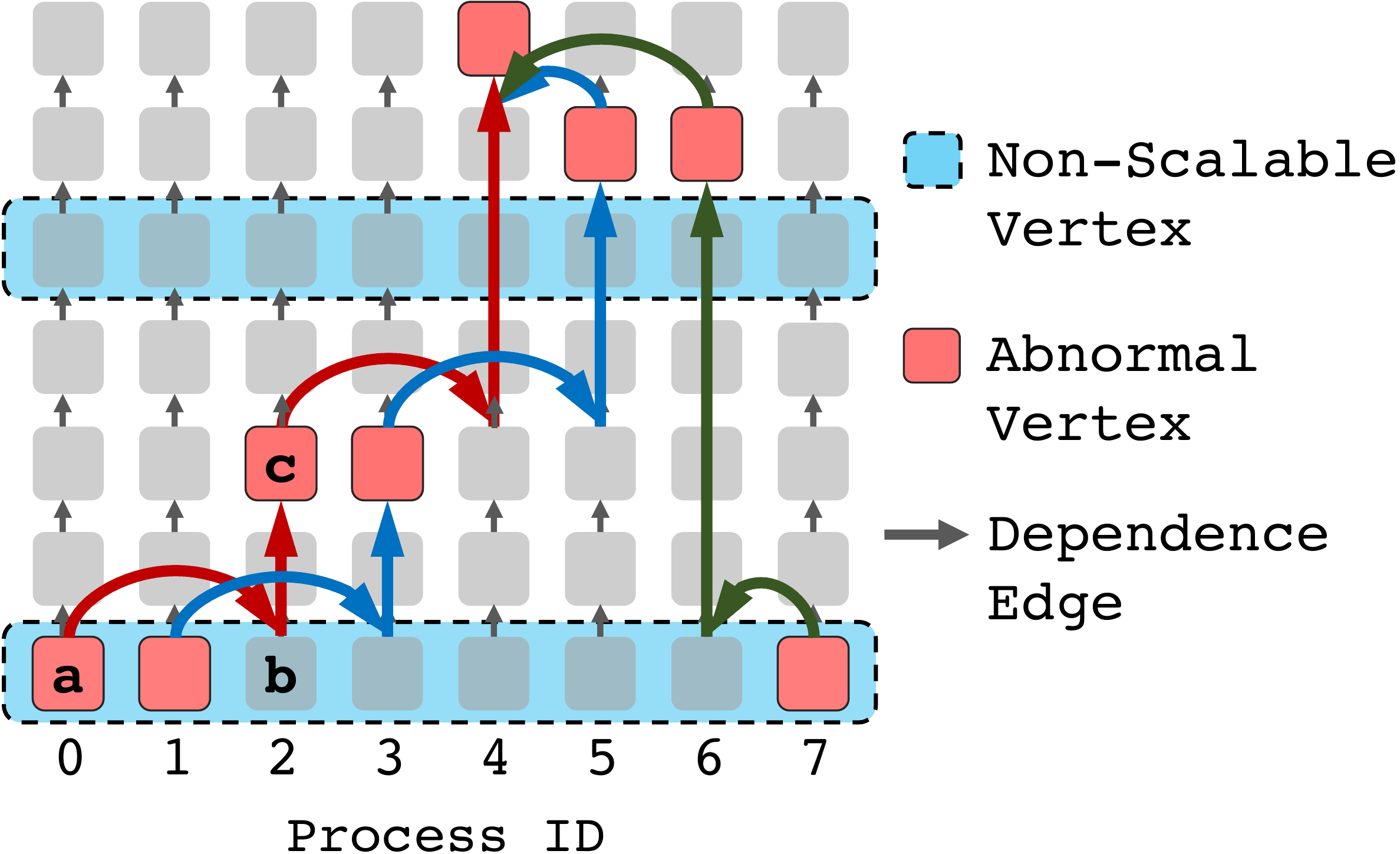}
	\caption{Problematic vertices and backtracking root cause detection on \ppg}
	\label{fig:cg-pagerank-result}
\end{figure}

For example, in Figure~\ref{fig:cg-pagerank-result}, we start from the \iv \textit{a} in the lower-left corner, and track through a communication dependence edge to vertex \textit{b} in process $2$. Then we can backtrack through the data dependence edge to vertex \textit{c} in process 2. We repeat the above steps and finally identify a path with the red color lines connecting all the \ivs in the processes of $0$, $2$, and $4$. With a similar approach, we backtrack from the other two \ivs, and then two extra paths are identified in Figure~\ref{fig:cg-pagerank-result}, shown as blue and green respectively. With these identified paths, we can connect different \ivs including MPI invocations and computation components together, and identify the root cause of scaling loss. 

Note that some vertices may be both non-scalable and \ivs. The interplay of non-scalable and \ivs can make the program performance even harder to understand. Sometimes, optimizing the performance of some vertices in identified paths can also improve the overall performance of the \nsv.

\section{Implementation and Usage}
For the static analysis module of \ourtool, we use LLVM-3.3.0 and Dragonegg-3.3.0~\cite{llvm} for \nomg generation and program instrumentation. For the sampling-based profiling, we use PAPI-5.2.0~\cite{papi,terpstra2010collecting} to collect hardware performance data, and the PMPI interface to collect communication dependencies. With both static \nomg and dynamic profiling data, \ourtool generates PPGs and performs scaling loss detection post-mortem.

In general, there are four main steps for end-users to use \ourtool :  (1) Compiling applications with \texttt{ScalAna-static} to generate the PSG.
(2) Running the instrumented applications with \texttt{ScalAna-prof} for different process numbers to collect profiling data. 
(3) Using \texttt{ScalAna-detect} to automatically detect the root cause of scaling loss.
(4) We also provide a GUI in \ourtool, \texttt{ScalAna-viewer}, to show the code snippets corresponding to the diagnosed root causes.
Besides, users can adjust some user-defined parameters like \texttt{MaxLoopDepth} and \texttt{AbnormThd} to make a trade-off between detection precision and system overhead.

\begin{figure}[h]
    \vspace{-1em}
	\centering
	\includegraphics[width=0.8\linewidth]{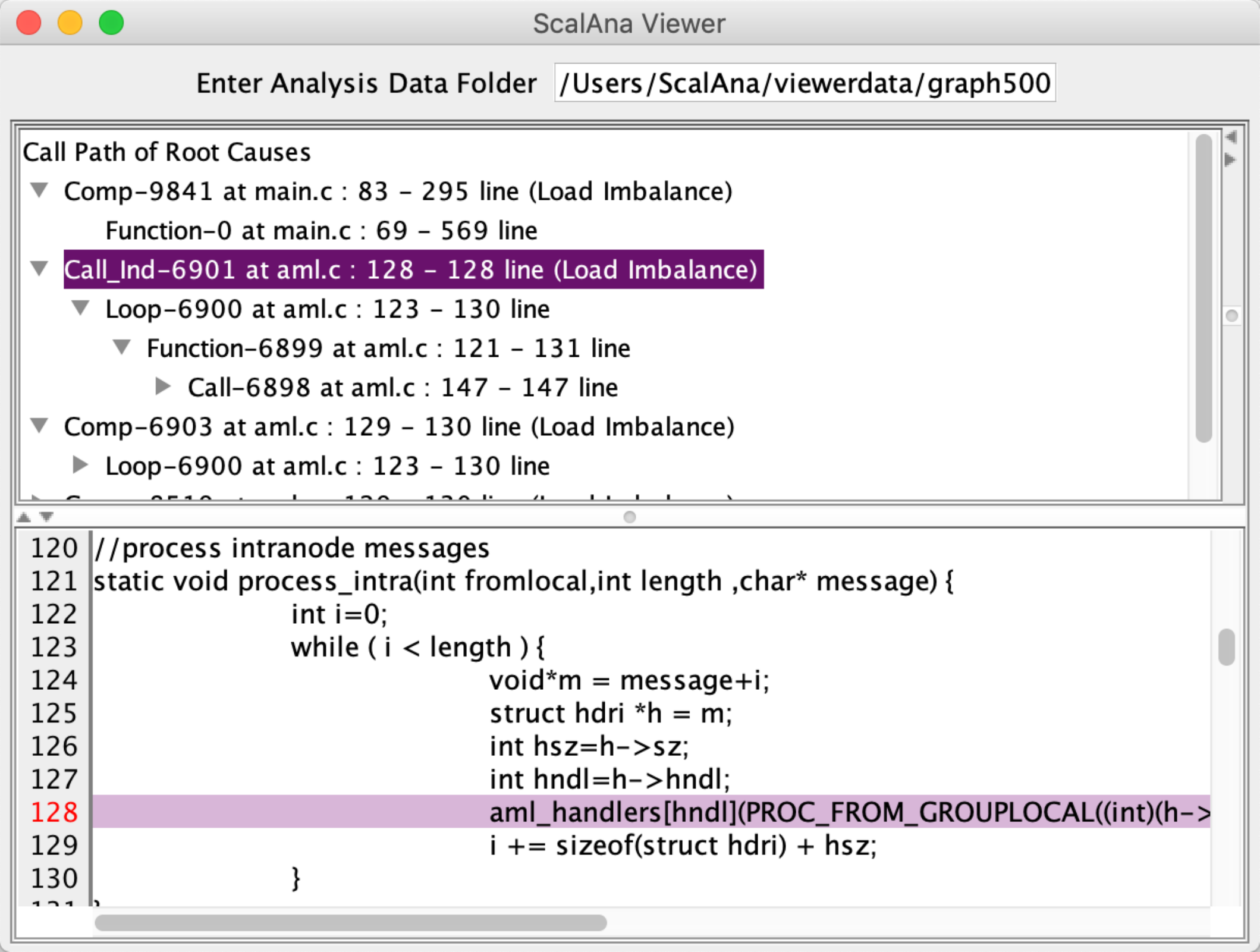}
	\caption{GUI of \ourtool}
	\label{fig:scalana-gui}
	\vspace{-1em}
\end{figure}

Figure~\ref{fig:scalana-gui} shows a screenshot of \ourtool's GUI. The upper window lists the root cause vertices and their calling paths. The lower window shows the code snippets corresponding to the vertices.
The root causes can be further sorted according to the length of execution time and the imbalance among different parallel processes. 

\ourtool currently only supports MPI-based programs in C or Fortran. However, all phases in \ourtool (program structure extraction, profiling data collection, and root-cause detection) are general enough to be adapted to other message-passing programs. In addition, our approach can be also extended to other programming models such as OpenMP or Pthreads with additional profiling techniques. We leave it for future work.
\section{Evaluation}
\label{sec:evalu}

\subsection{Experimental Setup}

\para{Experimental Platforms}
We perform the experiments on two testbeds: 
(1) Gorgon, a cluster with dual Intel Xeon E5-2670(v3) and 100Gbps 4xEDR Infiniband.
(2) Tianhe-2 supercomputer. Each node of Tianhe-2 has two Intel Xeon E5-2692(v2) processors (24 cores in total) and 64 GB memory. The Tianhe-2 supercomputer uses a customized high-speed interconnection network. 


\para{Evaluated Programs} We use a variety of parallel programs to evaluate the efficacy of \ourtool, including BT, CG, SP, EP, FT, MG, LU, and IS, from the widely used NPB benchmark suite~\cite{npb}, plus three real world applications, Zeus-MP~\cite{hayes2006simulating}, SST~\cite{rodrigues2011structural}, and Nekbone~\cite{fischer2008nek5000}.
For NPB programs, problem size CLASS C is used on Gorgon and CLASS D is used on the Tianhe-2 supercomputer.

\para{Methodology} In our evaluation, we first analyze the common features of the generated program structure graphs for each program, and then we present the performance overhead of our tool, including runtime overhead and storage cost (These experiments are for overhead comparison but not typical use-cases for scaling issue detection.). Finally, we use three real applications to demonstrate the benefit of our approach. For all experiments, we run three times and average the results for each process scale to reduce performance variance.

We compare our approach with two state-of-the-art performance tools, HPCToolkit~\cite{adhianto2010hpctoolkit} and Scalasca~\cite{geimer2010scalasca}. To ensure the fairness of comparison, we give the detailed configuration of these two state-of-the-art tools:
(1) For the tracing-based tool, Scalasca(v2.5), we first use its profiling function to identify where detailed tracing is needed, then we run small scale jobs with limited instrumentation. And increase the process count and the instrumentation complexity iteratively until the scalability bottlenecks are identified. In this way, Scalasca introduces as little storage cost as possible.
(2) For the profiling-based tool, HPCToolkit(v2019.08), the sampling frequency is the key parameter that affects the runtime overhead. \ourtool keeps the same sampling frequency (200Hz) as HPCToolkit in all experiments.
For all experiments, \texttt{MaxLoopDepth} is set to 10 and \texttt{AbnormThd} is set to 1.3 empirically.

\begin{table}[h]
\caption{Code size and vertices information of PSG for evaluated programs.
\#VBC and \#VAC are the number of vertices 
in the PSG before and after contraction, while \#\textit{Loop}, \#\textit{Branch}, \#\textit{Comp}, and \#\textit{MPI} are the number of \textit{Loop}, \textit{Branch}, \textit{Comp}, and \textit{MPI} vertices respectively.}
\label{tab:result}
\footnotesize
\begin{spacing}{0.9}
\begin{tabular}{p{1.3cm}|p{0.55cm} p{0.6cm}p{0.55cm}p{0.55cm}p{0.65cm}p{0.55cm}p{0.55cm}}
\hline
Program  & \begin{tabular}[c]{@{}c@{}}Code\\ (KLoc)\end{tabular} & \#VBC   & \#VAC  & \#\textit{Loop} & \#\textit{Branch} & \#\textit{Comp} & \#\textit{MPI} \\ \hline
BT       & 9.3                                                   & 974     & 377    & 39 & 57 & 176             & 103            \\
CG       & 2.0                                                   & 431     & 190    & 18 & 10 & 95              & 66             \\
EP       & 0.6                                                   & 91      & 32     & 4 & 2 & 13              & 12             \\
FT       & 2.5                                                   & 4,285   & 241    & 15 & 22 & 118             & 35             \\
MG       & 2.8                                                   & 7,842   & 1,973  & 177 & 233 & 942             & 463            \\
SP       & 5.1                                                   & 734     & 278    & 13 & 34 & 138             & 89             \\
LU       & 7.7                                                   & 2,370   & 663    & 18 & 66 & 327             & 237            \\
IS       & 1.3                                                   & 240     & 55     & 1 & 3 & 28              & 19             \\
SST      & 40.8                                                  & 23,608  & 5,217  & 321 & 641 & 1,434           & 1,303           \\
NEKBONE  & 31.8                                                  & 1,289   & 944    & 239 & 162 & 423             & 83             \\
ZEUS-MP  & 44.1                                                  & 273,715 & 64,570 & 1,677 & 1,304 & 30,099          & 11,818          \\
\hline
\end{tabular}
\end{spacing}
\end{table}

\subsection{PSG Analysis}

Table~\ref{tab:result} summarizes the code size and the vertices count for all generated PSGs. Results include the number of lines of source code, the number of vertices before and after graph contraction, the number of \textit{Comp} vertices, and the number of \textit{MPI} vertices.
In our experiments, the total vertex count correlates with the number of lines of source code in most cases.
Graph contraction reduces the number of vertices by 68\% on average.
Furthermore, \textit{Comp} and \textit{MPI} vertices make up more than 73\% of all vertices, which indicates that the \nomg can fully represent computation and communication characteristics.

\subsection{Performance Overhead}

We evaluate \ourtool on the Tianhe-2 supercomputer with up to 2,048 processes and the comparison experiments with Scalasca and HPCToolkit are run on Gorgon with up to 128 processes due to the installation limitation of the Tianhe-2 supercomputer's external network. 

\begin{table}[h!]
\caption{The static overhead of \ourtool on Gorgon}
\footnotesize
\begin{tabular}{p{1.0cm}|p{0.23cm}p{0.23cm}p{0.23cm}p{0.23cm}p{0.23cm}p{0.23cm}p{0.23cm}p{0.23cm}p{0.28cm}p{0.28cm}p{0.38cm}}
\toprule  
Programs&BT&CG&EP&FT&MG&SP&LU&IS&SST&NEK&ZMP\\
\midrule  
Ovd(\%)&0.32&0.77&0.38&0.35&0.29&0.31&0.28&0.68&3.01&0.43&2.96\\
\bottomrule 
\end{tabular}
\label{tab:static-cost}
\end{table}

\para{Static Overhead}
We first evaluate the compilation overhead introduced by static analysis on Gorgon. As shown in Table~\ref{tab:static-cost}, \ourtool only incurs very low compilation overhead comparing to the original LLVM compilation cost (0.28\% to 3.01\%, 0.89\% on average). 
Besides, the memory cost of static analysis is in proportion to the size of PSG. For example, each vertex of the PSG occupies 32B of memory on Gorgon and the static analysis incurs about 9MB in addition for Zeus-MP.

\begin{figure}[h!]
	\centering
    \includegraphics[width=0.9\linewidth]{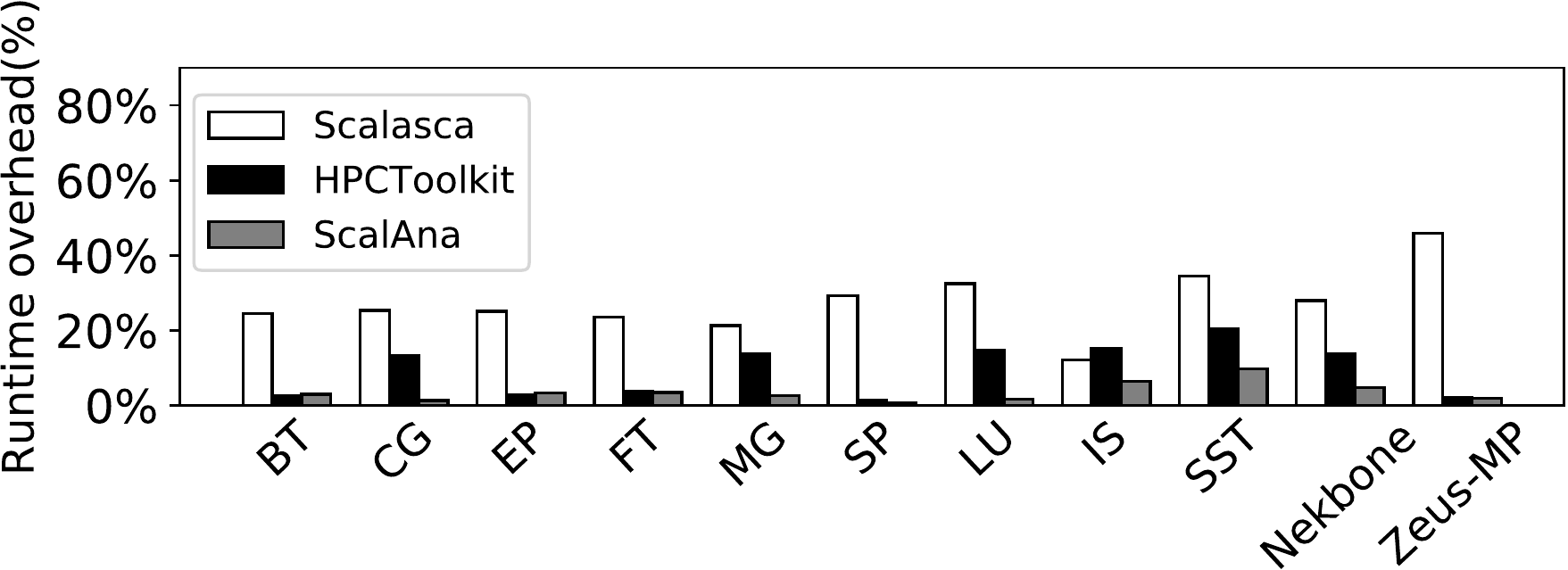}
    \vspace{-0.5em}
    \caption{Average runtime overhead of Scalasca~\cite{geimer2010scalasca}, HPCToolkit~\cite{adhianto2010hpctoolkit}, and \ourtool with 4 to 128 processes (without I/O)}
	\label{fig:time-ovh}
\end{figure}

\para{Runtime Overhead} 
The runtime overhead of \ourtool is shown as the gray bars in Figure~\ref{fig:time-ovh}, which is the average overhead of 4 to 128 MPI processes (4 to 121 processes for BT and SP, due to its requirement for process counts). 
As shown in Figure~\ref{fig:time-ovh}, \ourtool only brings very small overhead ranging from 0.72\% to 9.73\%, average at 3.52\% on Gorgon, which is much lower than Scalasca~\cite{geimer2010scalasca}. 
For Scalasca, the trace buffer size (\texttt{SCOREP\_TOTAL\_MEMORY}) is configured large enough to avoid intermediate trace flushing before the program ends.
Besides, for \ourtool, the average runtime overhead of the NPB benchmark with 2,048 processes on the Tianhe-2 supercomputer is 1.73\%.

\begin{figure}[h!]
	\centering
    \includegraphics[width=0.9\linewidth]{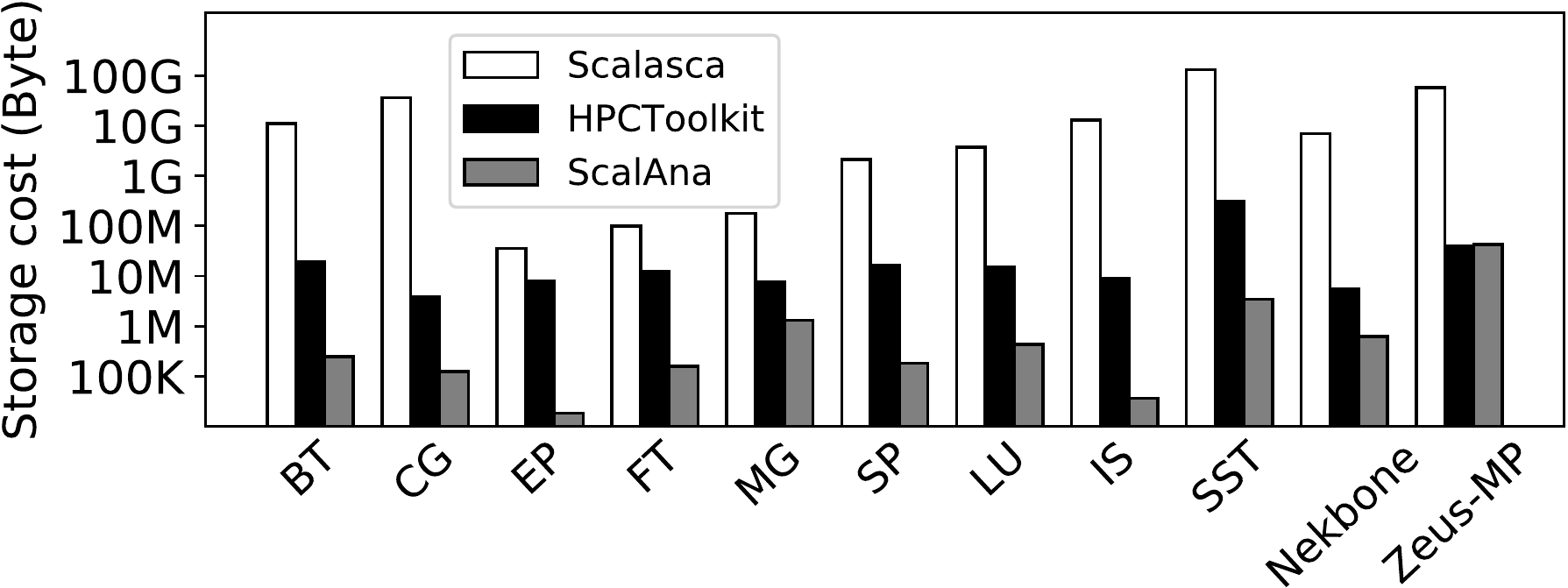}
    \vspace{-0.5em}
    \caption{Storage cost of Scalasca~\cite{geimer2010scalasca}, HPCToolkit~\cite{adhianto2010hpctoolkit}, and \ourtool running with 128 processes}
	\label{fig:storage-cost}
\end{figure}

\para{Storage Cost}
Figure~\ref{fig:storage-cost} shows the storage costs of \ourtool, HPCToolkit, and Scalasca running with 128 processes (121 for BT and SP) on Gorgon. \ourtool only incurs storage costs in the order of Kilobytes, while Scalasca and HPCToolkit generate Megabytes to Gigabytes of data.
Besides, for \ourtool, the average storage cost of the NPB benchmark with 2,048 processes on the Tianhe-2 supercomputer is 4.72MB.

\begin{table}[h!]
\caption{The post-mortem detection cost of \ourtool with 128 processes}
\footnotesize
\begin{tabular}{p{1.05cm}|p{0.23cm}p{0.23cm}p{0.23cm}p{0.23cm}p{0.23cm}p{0.23cm}p{0.23cm}p{0.23cm}p{0.28cm}p{0.28cm}p{0.38cm}}
\toprule  
Programs&BT&CG&EP&FT&MG&SP&LU&IS&SST&NEK&ZMP\\
\midrule  
Cost(Sec.)&3.26&1.74&0.29&2.20&1.80&2.40&6.06&0.50&9.54&8.63&11.81\\
\bottomrule 
\end{tabular}
\label{tab:post-cost}
\end{table}

\para{Post-mortem Detection Cost}
We evaluate the cost of backtracking root cause detection in \ourtool on Gorgon. As shown in Table~\ref{tab:post-cost}, the scaling loss detection only introduces little cost comparing to the execution time of the program (up to 11.81 seconds, 8.44\% of the execution time) on 128 processes. 
The memory consumption of post-mortem detection is proportional to the program structure and the size of profiling data (about 50MB for Zeus-MP on 128 processes).

\begin{figure*}[t]
	\centering
	\includegraphics[width=\textwidth]{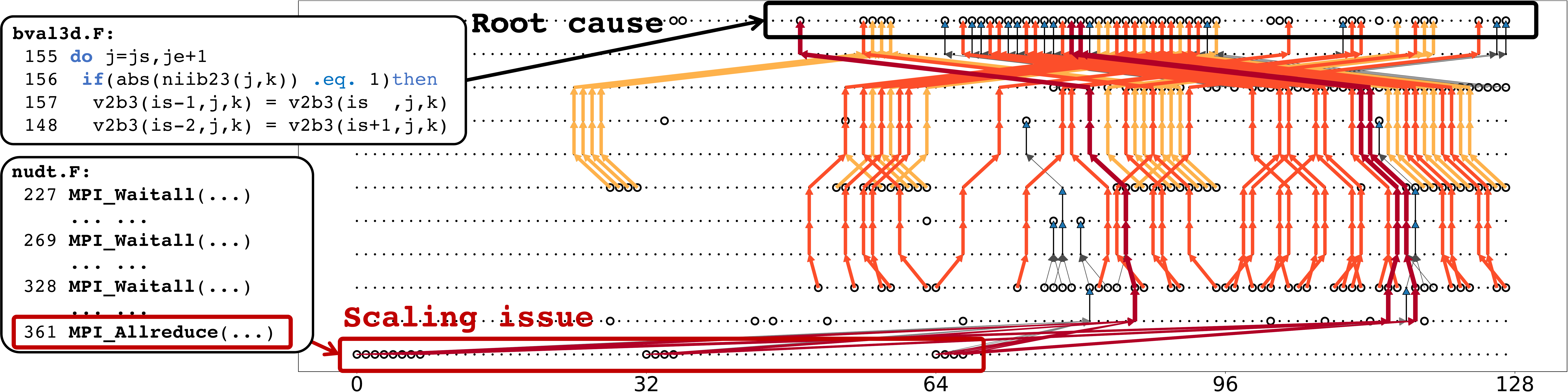}
	\vspace{-1.5em}
	\caption{Backtracking algorithm on the \ppg for a Zeus-MP run with 128 processes}
	\label{fig:zeusmp-backtracking}
\end{figure*}

\subsection{Case Studies with Real Applications}
\label{sub:eval-cases}
In this section, we use three real applications, Zeus-MP~\cite{hayes2006simulating}, SST~\cite{rodrigues2011structural}, and Nekbone~\cite{fischer2008nek5000}, to demonstrate how to diagnose scaling issues with our performance tool. When the root-causes of scaling issues are identified, we optimize the code to improve the scalability of these applications. We also analyze the advantages of our approach over the two state-of-the-art tools HPCToolkit~\cite{adhianto2010hpctoolkit} and Scalasca~\cite{geimer2010scalasca}.

\subsubsection{Zeus-MP}

Zeus-MP~\cite{hayes2006simulating}, a computational fluid dynamics program, implements the simulation of astrophysical phenomena in three spatial dimensions using the MPI programming model. Non-blocking point-to-point (P2P) communications are used to implement complex inter-process synchronization. We evaluate its performance with a problem size of 64$\times$64$\times$64 for different numbers of processes ranging from 4 to 128. We observe a significant scaling loss for 128 processes and results show that the speedup is only 55.53$\times$ on 128 processes while 35.40$\times$ on 64 processes (1 process as baseline). \ourtool is then applied to diagnose the problem.


\noindent \textbf{Scaling Loss Detection} \ourtool first generates a \ppg and then performs the backtracking algorithm on this graph to identify the root causes automatically. 
Figure~\ref{fig:zeusmp-backtracking} shows how \ourtool diagnoses the scaling issues on the \ppg of Zeus-MP by its backtracking algorithm. The vertical axis from top to down represents the control/data flow, and the horizontal axis represents different parallel processes. The small points represent the vertices of the \ppg with normal performance while the circle points represent problematic vertices with the abnormal performance for the same code snippets. The arrows show the backtracking paths based on intra- and inter-process dependence.

In detail, the \texttt{MPI\_Allreduce} at \textit{nudt.F: 361} is detected as a scaling issue due to its poor scalability for its execution time. 
As shown in Figure~\ref{fig:zeusmp-backtracking}, the dark red (darkest color) lines track backward from the abnormal \texttt{MPI\_Allreduce} vertices, then go through the intra-process dependence of control/data flow and inter-process dependence of P2P communications at \textit{nudt.F: 328, 269, 227}. 
The red (lighter color) and orange (lightest color) lines indicate similar backtracking paths. 
Finally, the \texttt{LOOP} vertices at \textit{bval3d.F: 155} (top row in Figure~\ref{fig:zeusmp-backtracking}) are identified as the root causes of scaling issues.

We find that the underlying reason is that only some busy processes execute the \texttt{LOOP} at \textit{bval3d.F: 155} while the others are idle with non-blocking P2P communications at \textit{nudt.F: 227}. 
Delays in these processes can propagate through the non-blocking P2P communications at \textit{nudt.F: 269} and \textit{nudt.F: 328}. The \texttt{MPI\_Allreduce} at \textit{nudt.F: 361} synchronizes all processes and leads to the low performance of Zeus-MP.

\noindent \textbf{Optimization} To fix the performance issue identified by \ourtool, 
we change the program into a hybrid programming model with MPI plus OpenMP, by adding multi-thread support at the \texttt{LOOP} of \textit{bval3d.F: 155}, which can accelerate the busy processes and mitigate the latent load imbalance between busy processes and idle processes.
Similarly, \ourtool also detects other root causes of the scaling loss from the \texttt{LOOP}s at \textit{hsmoc.F: 665, 841, 1,041}. 
\ourtool shows that the load/store instruction count and the cache miss count recorded by the PMU (Performance Monitor Unit) stays high with increasing numbers of processes. We use the techniques of loop tiling and scalar promotion to reduce the cache miss and memory access. 
With these optimizations, the speedup of Zeus-MP is increased from 55.53$\times$ to 61.39$\times$ (1 process as baseline) on 128 processes and a 9.55\% performance improvement is achieved on Gorgon.

We also test the optimized performance of Zeus-MP with a large process number. The speedup of Zeus-MP is increased from 68.41$\times$ to 76.15$\times$ (16 processes as baseline) on 2,048 processes and 9.96\% performance improvement is achieved on Tianhe-2 supercomputer. Note that more optimization techniques can be further explored for Zeus-MP, but we only give some common optimizations here to verify the performance bottlenecks detected by \ourtool.

\noindent \textbf{Comparison} As for other state-of-the-art tools, Scalasca can accurately detect the root causes at function-level when the number of processes increases to 64 with some human intervention. The profiling-based HPCToolkit can automatically detect the fine-grained loop-level scaling issues. Specifically, the \texttt{MPI\_Allreduce} at \textit{nudt.F: 361} and the \texttt{LOOP} at \textit{bval3d.F: 155} can be detected as scalability bottlenecks in HPCToolkit.
However, profiling-based HPCToolkit cannot easily identify the root cause problem (\texttt{LOOP} at \textit{bval3d.F: 155}) without significant human efforts. The outputs from HPCToolkit will show multiple bottlenecks without analysis on their underlying relationship to infer which one is the actual root cause.

\begin{figure}[h]
	\centering
	\begin{minipage}[t]{\linewidth}
	    \centering
		\subfigure[Runtime overhead]{
			\label{fig:zeusmp-run-ovhd}
			\includegraphics[width=0.46\linewidth]{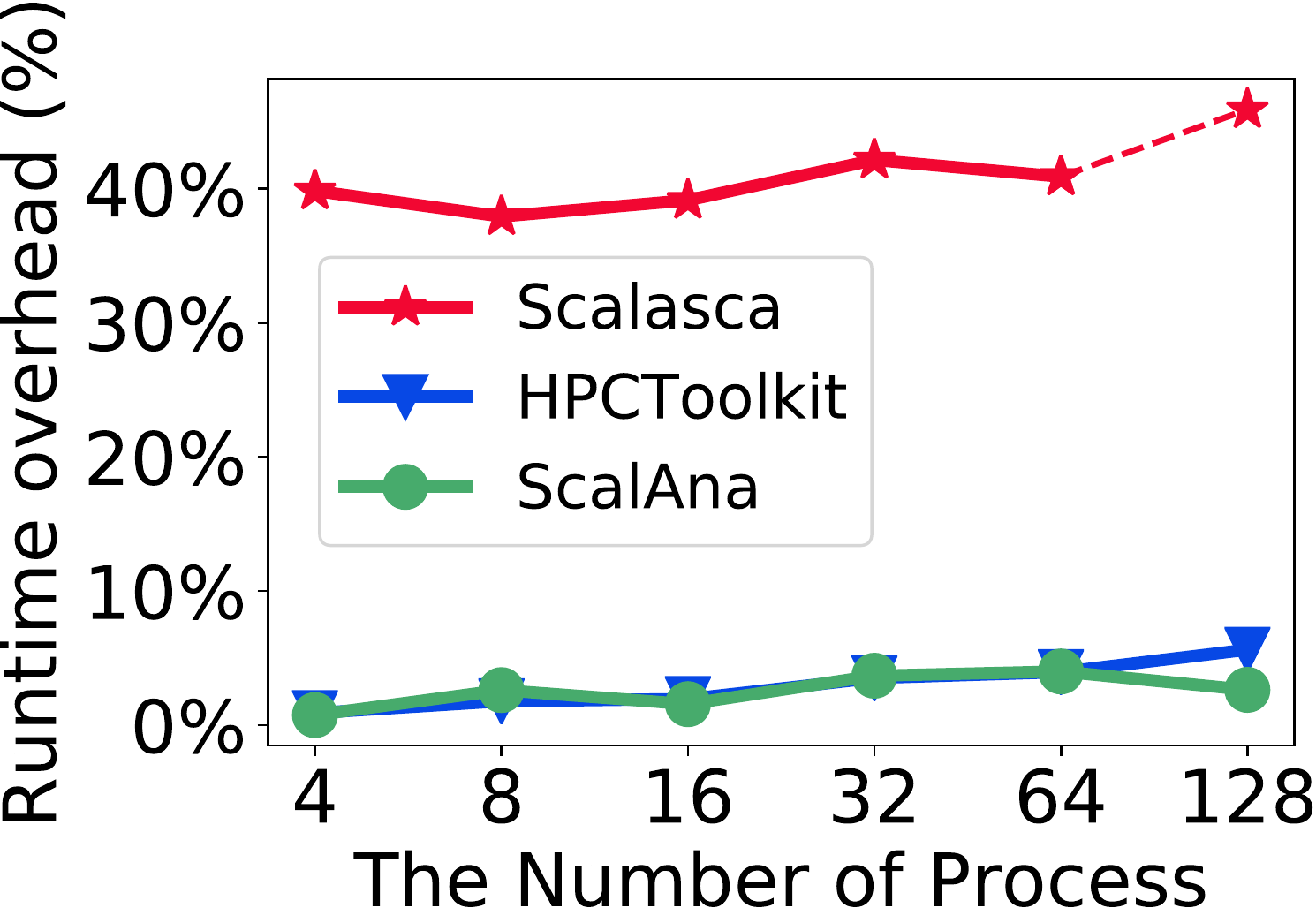} 
		}
		\subfigure[Storage cost]{
			\label{fig:zeusmp-spa-cost}
			\includegraphics[width=0.46\linewidth]{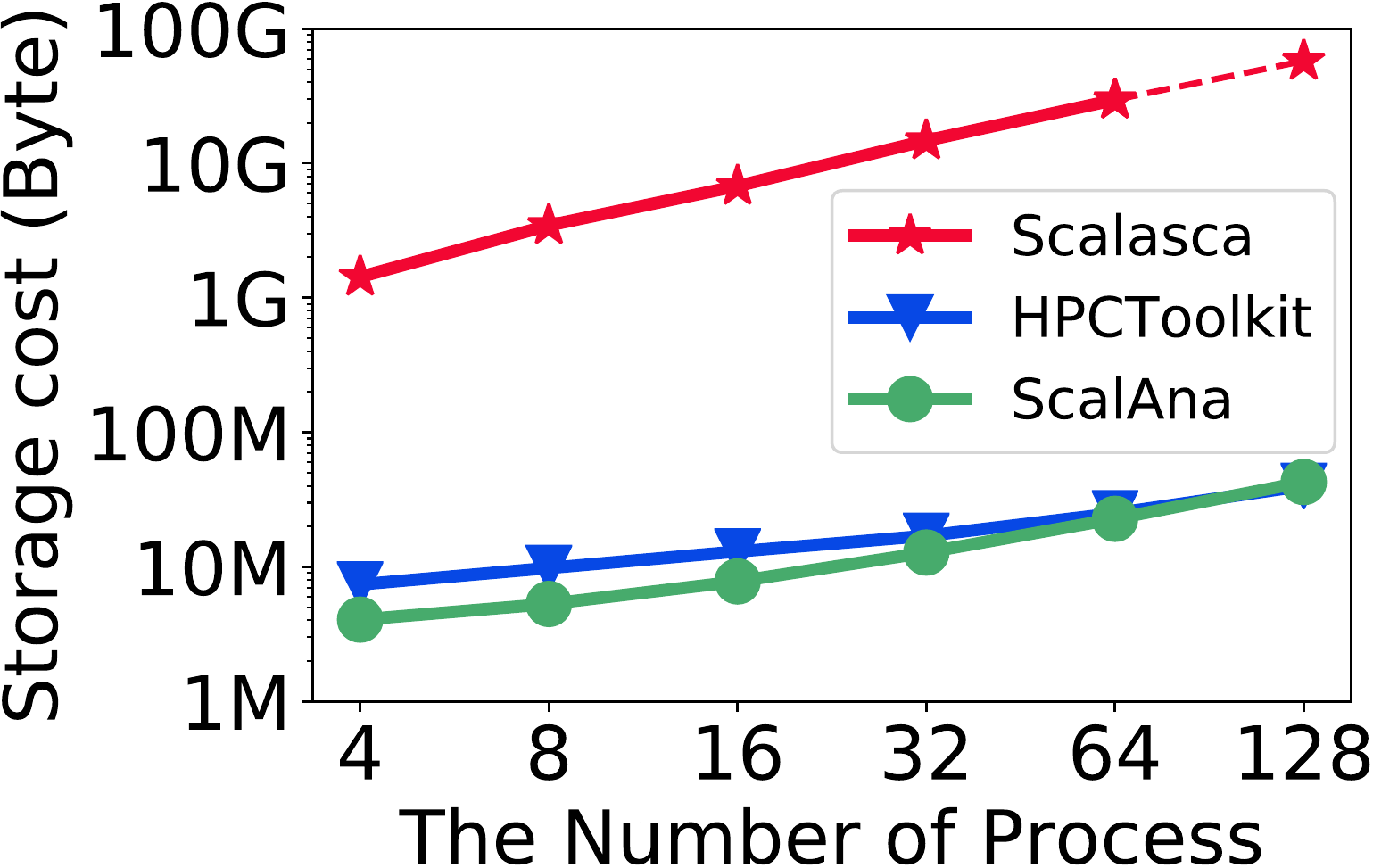} 
		}
	\end{minipage}
	\caption{Runtime and storage overhead of Scalasca~\cite{geimer2010scalasca}, HPCToolkit~\cite{adhianto2010hpctoolkit}, and \ourtool when running Zeus-MP (Scalasca detects the root cause when the number of processes increases to 64.)}
	\label{fig:zeusmp-overhead}
\end{figure}

Figure~\ref{fig:zeusmp-overhead} shows the performance and storage analysis of \ourtool against the state-of-the-art Scalasca and HPCToolkit. The lower is better for both Figure~\ref{fig:zeusmp-run-ovhd} and \ref{fig:zeusmp-spa-cost}. As for performance, both \ourtool and HPCToolkit have a negligible runtime overhead by 1.85\% and 2.01\% on average, respectively.  
However, the tracing-based Scalasca introduces 40.89\% runtime overhead on 64 processes (without I/O) to generate traces. 
For storage, our light-weight \ourtool is better than Scalasca. \ourtool only needs 20MB storage space while Scalasca generates 28.26GB traces of 64 processes.


\subsubsection{SST}
SST (Structural Simulation Toolkit)~\cite{rodrigues2011structural}
is a multi-process simulation framework, which simulates for microarchitecture and memory in highly concurrent systems. 
We execute SST for different numbers of processes ranging from 4 to 128, 
and results show that the speedup is only 1.20$\times$ on 32 processes while 1.28$\times$ on 16 processes (4 processes as baseline). We notice that the dependence of simulated events in SST is usually complex so that most events need to be executed sequentially. The parallelism only occurs within each event in most cases, causing relatively low speedup for 32 processes.
We use \ourtool to analyze the scaling loss of SST. 

\begin{figure}[h!]
	\centering
	\includegraphics[width=\linewidth]{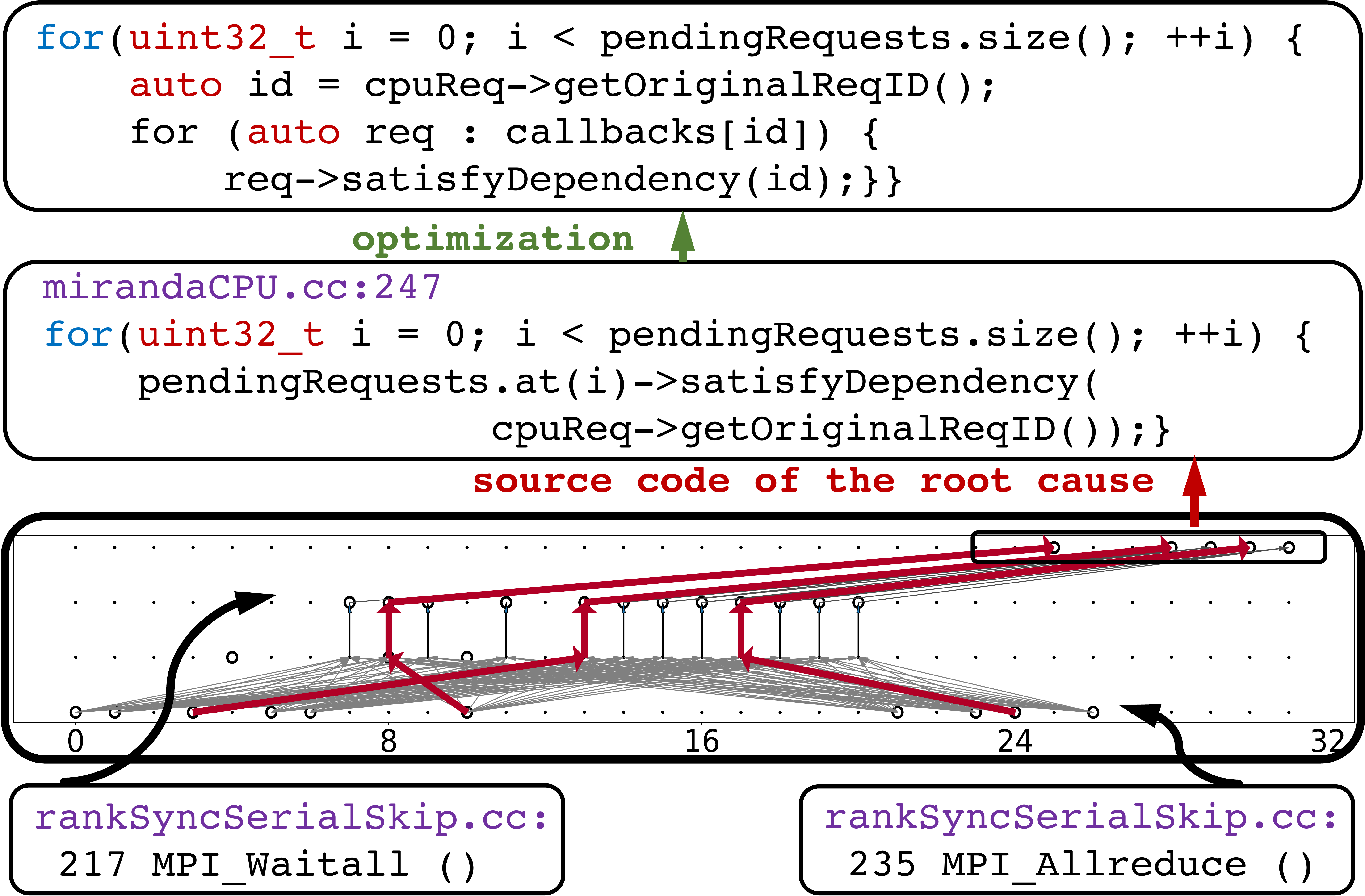}
	\caption{Backtracking algorithm on \ppg and code optimization for an SST run with 32 processes}
	\label{fig:sst-backtracing}
\end{figure}

\noindent \textbf{Scaling Loss Detection} \ourtool finds that the scaling loss mainly comes from the \texttt{MPI\_Allreduce} in the \textit{RankSyncSerialSkip::exchange} function at \textit{rankSyncSerialSkip.cc:235}.
As shown in Figure~\ref{fig:sst-backtracing}, after backward tracking through P2P communications \texttt{MPI\_Waitall} in the function \textit{RankSyncSerialSkip::exchange} at \textit{rankSyncSerialSkip.cc:217}, 
the \texttt{LOOP} in the function \textit{RequestGenCPU::handleEvent} at \textit{mirandaCPU.cc:247} 
is identified as the root cause of scaling issues. The colored lines show some backtracking paths as examples.
\begin{figure}[h!]
	\centering
	\includegraphics[width=\linewidth]{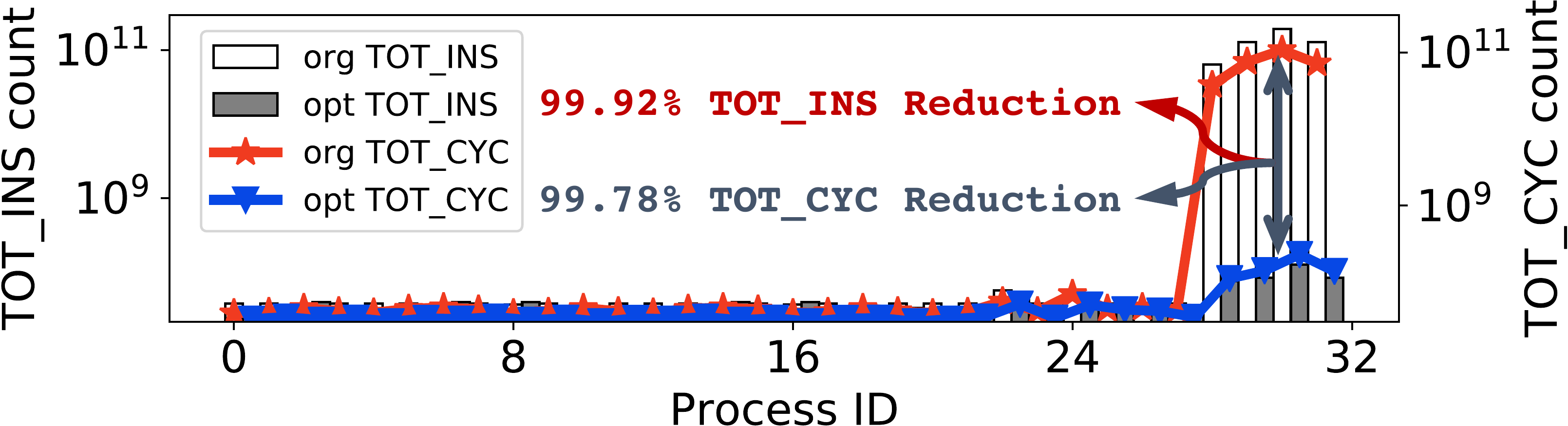}
	\vspace{-1em}
	\caption{PMU data for SST running with 32 processes}
	\label{fig:sst-pmu}
\end{figure}

\noindent \textbf{Optimization} As shown in Figure~\ref{fig:sst-pmu}, \ourtool provides the PMU data showing that the total instruction counts (\texttt{TOT\_INS}) for different processes differ a lot in this loop. 
Based on the results of \ourtool, we find that this program uses an inefficient data structure (\textit{array}) to process each query in a critical path for each process, which can cause different execution time (\texttt{TOT\_INS}) to traverse the array for different processes.
We modify the code and change the data structure from \textit{array} to \textit{unordered\_map}, which reduces the complexity of the query algorithm from \textit{O(n)} to \textit{O($\log$(n))} and makes the load (execution time of query) of different processes more balanced.
Figure~\ref{fig:sst-pmu} also shows \texttt{TOT\_INS} counts of different processes after our optimization, which are more balanced among different processes than the original SST.
After the optimization, the speedup of SST for 32 processes is increased from 1.20$\times$ to 1.56$\times$ (4 processes as baseline) and the performance is improved by 73.12\%.

\noindent \textbf{Comparison} The state-of-the-art profiling tool HPCToolkit only locates that \texttt{MPI\_Waitall} is a scalability bottleneck but not the \texttt{LOOP} in the function \textit{RequestGenCPU::handleEvent} because it does not profile on threads created at runtime, although its method is able to profile the threads theoretically.
Even if it can do profiling on threads, the root cause identification still needs more human analysis.
Besides, \ourtool provides the PMU data of the root causes, which makes it possible to analyze on an architecture level for developers.
For storage, \ourtool only needs 1.03MB storage space while Scalasca needs 31.56GB to store the generated traces of 32 processes.

\subsubsection{Nekbone}
Nekbone, the basic structure of Nek5000~\cite{fischer2008nek5000}, uses a spectral element method to solve the Helmholtz equation in three-dimensional space. We execute Nekbone at the scale of 16,384 elements for the number of processes ranging from 4 to 128. 
Nekbone encounters a scaling issue when running on 64 processes. The speedup is only 31.95$\times$ for 64 processes while the speedup of 32 processes is 20.61$\times$ (1 process as baseline).

\noindent \textbf{Scaling Loss Detection} We use \ourtool to analyze the root cause of the scalability problem.
\texttt{MPI\_Waitall} in the function of \textit{comm\_wait} at \textit{comm.h:243} is detected as a \nsv.
Using the backtracking algorithm on the \ppg through inter-process dependence, \ourtool finds that the root cause of the scaling loss is the \texttt{LOOP} in the function of \textit{dgemm} at \textit{blas.f:8,941}. 
In this loop, some processes consume significantly less time than others, which causes the waiting time of \texttt{MPI\_Waitall} to increase and finally leads to the poor scalability of Nekbone.

\begin{figure}[h!]
	\centering
	\includegraphics[width=\linewidth]{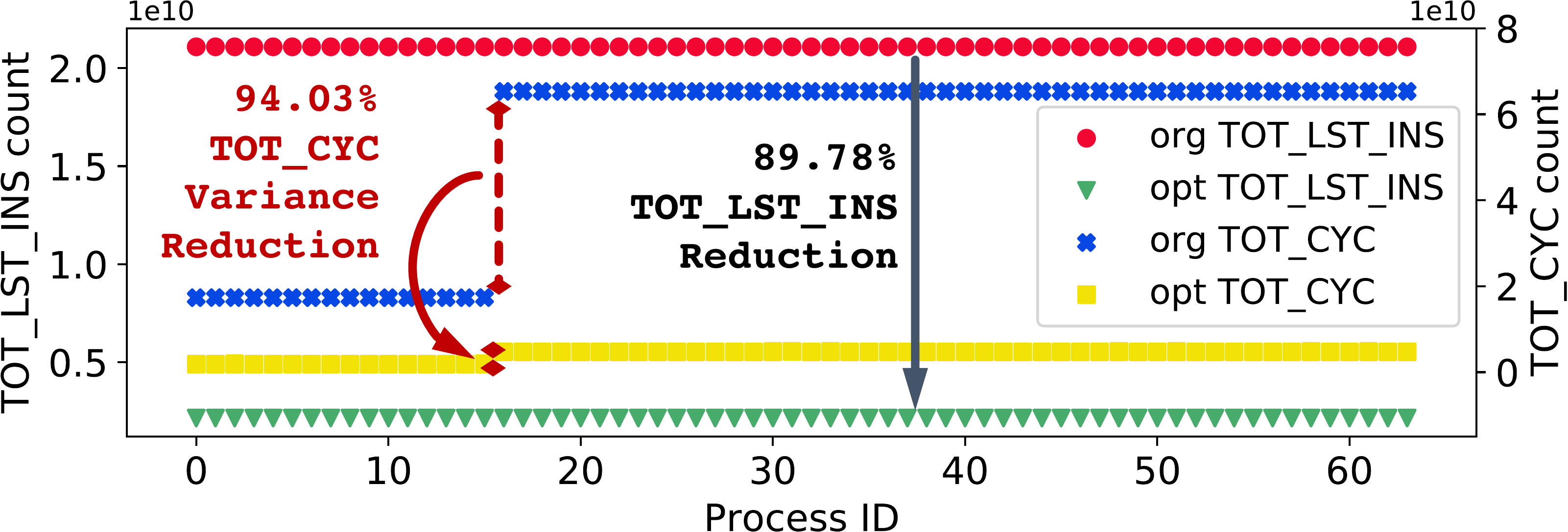}
	\caption{PMU data for Nekbone running with 32 processes}
	\label{fig:nek-pmu}
\end{figure}

\noindent \textbf{Optimization} As shown in Figure~\ref{fig:nek-pmu}, the PMU data provided by \ourtool shows that the load/store instruction count (\texttt{TOT\_LST\_INS}) of this loop is the same among processes while the total cycle count (\texttt{TOT\_CYC}) of the loop differs. We find that the memory access speed of each processor core differs, and the processes are bound to different processor cores.
From the perspective of the code, we optimize it by using a more efficient linear algebra library (BLAS) to reduce the number of \texttt{TOT\_LST\_INS} and mitigate the time variance among processes.
Figure~\ref{fig:nek-pmu} also shows that \texttt{TOT\_LST\_INS} decreases by 89.78\%, and the execution time variance among different processes is reduced by 94.03\%. 
After the optimization, the speedup on 64 processes is increased from 31.95$\times$ to 51.96$\times$ (1 process as baseline) and the performance is improved by 68.95\%.

We also analyze the optimized performance of Nekbone with a large process number. The speedup on 2,048 processes is increased from 27.08$\times$ to 29.97$\times$ (64 processes as baseline) and 11.11\% performance improvement is achieved on Tianhe-2 supercomputer.

\noindent \textbf{Comparison} 
For HPCToolkit, the \texttt{MPI\_Waitall} at \textit{comm.h: 243}, the \texttt{LOOP} at \textit{blas.f: 8,941}, and some other points are detected as potential bottlenecks, but further manual analysis is needed to find the root cause.
For storage, \ourtool only needs 0.32MB storage space while Scalasca needs 3.44GB to store the generated traces of 64 processes.

\section{Related Work}
\label{sec:related}
Mohr~\cite{mohr2014scalable} gives a comprehensive survey of state-of-the-art performance analysis tools including both tracing- and profiling-based methods. Knobloch et al.~\cite{knobloch2020tools} present a sufficient survey of performance tools for heterogeneous HPC applications. In the remaining part of this section, we discuss representative related work for performance analysis in detail.

\para{Tracing}
Traces are widely used for analyzing program behavior. 
Intel provides a trace collection tool to understand MPI program's behavior~\cite{itac}.
Based on Score-P infrastructure~\cite{an2011score,scorephp}, TAU~\cite{shende2006tau,tauhp}, Vampir~\cite{nagel1996vampir,muller2007developing,knupfer2008vampir,vampirhg}, Scalasca~\cite{geimer2010scalasca,scalascahg}, and some state-of-the-art tools support various programming models, such as MPI, OpenMP, Pthread, and CUDA. These tools can visualize trace data and provide fine-grained performance analysis for developers.
Paraver~\cite{labarta1996dip,servat2009detailed,paraver} is a tracing-based performance analyzer that supports flexible data collection and detailed analysis of metrics variability. 
Becker et al.~\cite{becker2007automatic} use event traces to analyze the performance for large-scale programs. Though many works for trace compression are proposed~\cite{zhai2014cypress,noeth2007scalable,krishnamoorthy2010scalable,knupfer2005construction}, 
tracing still often brings very large overhead which makes it non-suitable for production environments.

\para{Profiling}
Profiling can extract the program's statistical information with very low overhead. mpiP~\cite{vetter2005mpip} is a light-weight profiling library for MPI applications, 
which can collect statistical information for MPI functions with low overhead.
Tallent et al.~\cite{tallent2010scalable,tallent2009diagnosing} use call path profiling to identify and qualify the load imbalance for parallel programs.
STAT~\cite{arnold2007stack} performs large scale debugging by sampling stack trace to assemble a profile for applications' behavior.
HPCToolkit~\cite{adhianto2010hpctoolkit} uses sample-based techniques to get the profile performance of applications and visualize the results with \textit{hpcviewer} and \textit{hpctraceviewer}.
Arm MAP~\cite{january2015allinea} is a light-weight profiler, which is available as a part of Arm Forge debug and profile suite.
Cray develops CrayPat~\cite{kaufmann2003craypat}, supporting both tracing and profiling performance analysis, for XC platforms. 
However, profiling often misses important information which may prevent us from correctly understanding the program's behavior. 

Our approach uses profiling to collect dynamic statistical information,
while combining it with static extracted program structure, so that we can achieve high accuracy with low overhead.

\para{Program structure based program analysis}
Cypress~\cite{zhai2014cypress} and Spindle~\cite{wang2018spindle} use hybrid static-dynamic analysis for communication trace compression and memory access monitoring.
By extracting the program structure at compilation time, the runtime overhead can be significantly reduced.
Weber et al.~\cite{weber2016structural} presents effective structural
similarity measure to classify and store the data for parallel programs.
Program structure is also used for large scale debugging~\cite{laguna2015debugging,zhou2011vrisha,laguna2011large,mitra2014accurate},
since program structure contains the dependence for both inter- and intra-process, which play an important role in large scale debugging.

\para{Detecting scalability bottlenecks}
Coarfa et al.~\cite{coarfa2007scalability} identify the scalability bottlenecks by analyzing 
HPCToolkit's~\cite{adhianto2010hpctoolkit} hpcviewer data with a top down approach. 
However, it cannot deal with some communication patterns with complex dependence.
Bohme et al.~\cite{bohme2010identifying} use runtime trace to identify the root cause of wait states. 
As a tracing-based approach, Bohme's work performs a forward and backward trace replay on collected timeline traces. 
With the complete traces, delay or root causes can be accurately identified.
Inspired by Bohme's backward-replay analysis, we propose a backtracking root cause detection algorithm in \ourtool. Instead of recording a large amount of traces, our approach works on the program structure based \ppg, which contains little profiling data. Therefore, \ourtool introduces very low storage cost and detection overhead.
Barnes et al.~\cite{barnes2008regression} use regression-based approaches 
to perform scalability prediction. 
Calotoiu et al.~\cite{calotoiu2013using} automate traditional performance modeling to detect scalability bugs. Bhattacharyya et al.~\cite{bhattacharyya2014pemogen} improve it using compiler techniques. 
Chen et al.~\cite{chen2015critical} present a scalable performance modeling 
framework based on the concept of critical-path candidates for MPI workloads.
ScaAnalyzer~\cite{liu2015scaanalyzer} collects, attributes, and analyzes memory-related metrics at runtime to identify the scalability bottlenecks caused by memory access behavior for the parallel programs running on a single node. 
COLAB~\cite{yu2020colab} collects and accumulates futexes from Linux kernel at runtime to detect bottlenecks caused by program synchronizations.

Our work targets on detecting scalability bottlenecks using program structure combining with runtime profiling information, which helps address the root cause more accurately.

\section{Conclusion}
\label{sec:con}
In this paper, we design \ourtool, a light-weight performance tool that can efficiently detect scalability problems of parallel programs by combining both static and dynamic analysis. \ourtool uses a novel approach to automatically identify the root cause for complex parallel programs, named backtracking root cause detection, through traversing a program performance graph. We evaluate it with both benchmarks and applications. Results show that \ourtool can efficiently identify the scalability bottlenecks with very low overhead and outperform state-of-the-art approaches.




\bibliographystyle{IEEEtran}
\bibliography{refs}

\begin{thebibliography}{10}
\providecommand{\url}[1]{#1}
\csname url@samestyle\endcsname
\providecommand{\newblock}{\relax}
\providecommand{\bibinfo}[2]{#2}
\providecommand{\BIBentrySTDinterwordspacing}{\spaceskip=0pt\relax}
\providecommand{\BIBentryALTinterwordstretchfactor}{4}
\providecommand{\BIBentryALTinterwordspacing}{\spaceskip=\fontdimen2\font plus
\BIBentryALTinterwordstretchfactor\fontdimen3\font minus
  \fontdimen4\font\relax}
\providecommand{\BIBforeignlanguage}[2]{{%
\expandafter\ifx\csname l@#1\endcsname\relax
\typeout{** WARNING: IEEEtran.bst: No hyphenation pattern has been}%
\typeout{** loaded for the language `#1'. Using the pattern for}%
\typeout{** the default language instead.}%
\else
\language=\csname l@#1\endcsname
\fi
#2}}
\providecommand{\BIBdecl}{\relax}
\BIBdecl

\bibitem{top500}
\BIBentryALTinterwordspacing
``top500 website,'' 2020. [Online]. Available: \url{http://top500.org/}
\BIBentrySTDinterwordspacing

\bibitem{shi2012program}
J.~Y. Shi, M.~Taifi, A.~Pradeep, A.~Khreishah, and V.~Antony, ``Program
  scalability analysis for hpc cloud: Applying amdahl's law to nas
  benchmarks,'' in \emph{2012 SC Companion: High Performance Computing,
  Networking Storage and Analysis}.\hskip 1em plus 0.5em minus 0.4em\relax
  IEEE, 2012, pp. 1215--1225.

\bibitem{liu2015scaanalyzer}
X.~Liu and B.~Wu, ``Scaanalyzer: A tool to identify memory scalability
  bottlenecks in parallel programs,'' in \emph{Proceedings of the International
  Conference for High Performance Computing, Networking, Storage and
  Analysis}.\hskip 1em plus 0.5em minus 0.4em\relax ACM, 2015, p.~47.

\bibitem{pearce2019exploring}
O.~Pearce, H.~Ahmed, R.~W. Larsen, P.~Pirkelbauer, and D.~F. Richards,
  ``Exploring dynamic load imbalance solutions with the comd proxy
  application,'' \emph{Future Generation Computer Systems}, vol.~92, pp.
  920--932, 2019.

\bibitem{schmidl2016openmp}
D.~Schmidl, M.~S. M{\"u}ller, and C.~Bischof, ``Openmp scalability limits on
  large smps and how to extend them,'' Fachgruppe Informatik, Tech. Rep., 2016.

\bibitem{npb}
D.~Bailey, T.~Harris, W.~Saphir, R.~V.~D. Wijngaart, A.~Woo, and M.~Yarrow,
  \emph{The {NAS} Parallel Benchmarks 2.0}.\hskip 1em plus 0.5em minus
  0.4em\relax Moffett Field, CA: NAS Systems Division, NASA Ames Research
  Center, 1995.

\bibitem{geimer2010scalasca}
M.~Geimer, F.~Wolf, B.~J. Wylie, E.~{\'A}brah{\'a}m, D.~Becker, and B.~Mohr,
  ``The scalasca performance toolset architecture,'' \emph{Concurrency and
  Computation: Practice and Experience}, vol.~22, no.~6, pp. 702--719, 2010.

\bibitem{adhianto2010hpctoolkit}
L.~Adhianto, S.~Banerjee, M.~Fagan, M.~Krentel, G.~Marin, J.~Mellor-Crummey,
  and N.~R. Tallent, ``Hpctoolkit: Tools for performance analysis of optimized
  parallel programs,'' \emph{Concurrency and Computation: Practice and
  Experience}, vol.~22, no.~6, pp. 685--701, 2010.

\bibitem{vetter2005mpip}
J.~Vetter and C.~Chambreau, ``mpip: Lightweight, scalable mpi profiling,''
  2005.

\bibitem{tallent2010scalable}
N.~R. Tallent, L.~Adhianto, and J.~M. Mellor-Crummey, ``Scalable identification
  of load imbalance in parallel executions using call path profiles,'' in
  \emph{Proceedings of the 2010 ACM/IEEE International Conference for High
  Performance Computing, Networking, Storage and Analysis}.\hskip 1em plus
  0.5em minus 0.4em\relax IEEE Computer Society, 2010, pp. 1--11.

\bibitem{tallent2009diagnosing}
N.~R. Tallent, J.~M. Mellor-Crummey, L.~Adhianto, M.~W. Fagan, and M.~Krentel,
  ``Diagnosing performance bottlenecks in emerging petascale applications,'' in
  \emph{Proceedings of the Conference on High Performance Computing Networking,
  Storage and Analysis}.\hskip 1em plus 0.5em minus 0.4em\relax IEEE, 2009, pp.
  1--11.

\bibitem{itac}
\BIBentryALTinterwordspacing
``Intel trace analyzer and collector.'' [Online]. Available:
  \url{https://software.intel.com/en-us/trace-analyzer}
\BIBentrySTDinterwordspacing

\bibitem{zhai2010phantom}
J.~Zhai, W.~Chen, and W.~Zheng, ``Phantom: predicting performance of parallel
  applications on large-scale parallel machines using a single node,'' in
  \emph{ACM Sigplan Notices}, vol.~45, no.~5.\hskip 1em plus 0.5em minus
  0.4em\relax ACM, 2010, pp. 305--314.

\bibitem{linford2017performance}
J.~C. Linford, S.~Khuvis, S.~Shende, A.~Malony, N.~Imam, and M.~G. Venkata,
  ``Performance analysis of openshmem applications with tau commander,'' in
  \emph{Workshop on OpenSHMEM and Related Technologies}.\hskip 1em plus 0.5em
  minus 0.4em\relax Springer, 2017, pp. 161--179.

\bibitem{yin2016discovering}
H.~Yin, Z.~Hu, X.~Zhou, H.~Wang, K.~Zheng, Q.~V.~H. Nguyen, and S.~Sadiq,
  ``Discovering interpretable geo-social communities for user behavior
  prediction,'' in \emph{2016 IEEE 32nd International Conference on Data
  Engineering}.\hskip 1em plus 0.5em minus 0.4em\relax IEEE, 2016, pp.
  942--953.

\bibitem{yin2016joint}
H.~Yin, B.~Cui, X.~Zhou, W.~Wang, Z.~Huang, and S.~Sadiq, ``Joint modeling of
  user check-in behaviors for real-time point-of-interest recommendation,''
  \emph{ACM Transactions on Information Systems}, vol.~35, no.~2, p.~11, 2016.

\bibitem{bhattacharyya2014pemogen}
A.~Bhattacharyya, G.~Kwasniewski, and T.~Hoefler, ``{Using Compiler Techniques
  to Improve Automatic Performance Modeling}.''\hskip 1em plus 0.5em minus
  0.4em\relax ACM, Oct. 2015, in proceedings of the 24th International
  Conference on Parallel Architectures and Compilation.

\bibitem{calotoiu2013using}
A.~Calotoiu, T.~Hoefler, M.~Poke, and F.~Wolf, ``Using automated performance
  modeling to find scalability bugs in complex codes,'' in \emph{Proceedings of
  the International Conference on High Performance Computing, Networking,
  Storage and Analysis}.\hskip 1em plus 0.5em minus 0.4em\relax ACM, 2013,
  p.~45.

\bibitem{wolf2016automatic}
F.~Wolf, C.~Bischof, A.~Calotoiu, T.~Hoefler, C.~Iwainsky, G.~Kwasniewski,
  B.~Mohr, S.~Shudler, A.~Strube, A.~Vogel \emph{et~al.}, ``Automatic
  performance modeling of hpc applications,'' in \emph{Software for Exascale
  Computing-SPPEXA 2013-2015}.\hskip 1em plus 0.5em minus 0.4em\relax Springer,
  2016, pp. 445--465.

\bibitem{beckingsale2017apollo}
D.~Beckingsale, O.~Pearce, I.~Laguna, and T.~Gamblin, ``Apollo: Reusable models
  for fast, dynamic tuning of input-dependent code,'' in \emph{2017 IEEE
  International Parallel and Distributed Processing Symposium (IPDPS)}.\hskip
  1em plus 0.5em minus 0.4em\relax IEEE, 2017, pp. 307--316.

\bibitem{linford2009multi}
J.~C. Linford, J.~Michalakes, M.~Vachharajani, and A.~Sandu, ``Multi-core
  acceleration of chemical kinetics for simulation and prediction,'' in
  \emph{Proceedings of the Conference on High Performance Computing Networking,
  Storage and Analysis}, 2009, pp. 1--11.

\bibitem{7776507}
A.~{Calotoiu}, D.~{Beckinsale}, C.~W. {Earl}, T.~{Hoefler}, I.~{Karlin},
  M.~{Schulz}, and F.~{Wolf}, ``Fast multi-parameter performance modeling,'' in
  \emph{2016 IEEE International Conference on Cluster Computing}, Sep. 2016,
  pp. 172--181.

\bibitem{llvm}
\BIBentryALTinterwordspacing
``The {LLVM} compiler framework.'' [Online]. Available: \url{http://llvm.org}
\BIBentrySTDinterwordspacing

\bibitem{nagel1996vampir}
W.~E. Nagel, A.~Arnold, M.~Weber, H.-C. Hoppe, and K.~Solchenbach, ``Vampir:
  Visualization and analysis of mpi resources,'' 1996.

\bibitem{papi}
\BIBentryALTinterwordspacing
``{PAPI} tools.'' [Online]. Available: \url{http://icl.utk.edu/papi/software/}
\BIBentrySTDinterwordspacing

\bibitem{wu2011scalaextrap}
X.~Wu and F.~Mueller, ``Scalaextrap: Trace-based communication extrapolation
  for spmd programs,'' in \emph{ACM SIGPLAN Notices}, vol.~46, no.~8.\hskip 1em
  plus 0.5em minus 0.4em\relax ACM, 2011, pp. 113--122.

\bibitem{noeth2009scalatrace}
M.~Noeth, P.~Ratn, F.~Mueller, M.~Schulz, and B.~R. De~Supinski, ``Scalatrace:
  Scalable compression and replay of communication traces for high-performance
  computing,'' \emph{Journal of Parallel and Distributed Computing}, vol.~69,
  no.~8, pp. 696--710, 2009.

\bibitem{vetter2002dynamic}
J.~Vetter, ``Dynamic statistical profiling of communication activity in
  distributed applications,'' \emph{ACM SIGMETRICS Performance Evaluation
  Review}, vol.~30, no.~1, pp. 240--250, 2002.

\bibitem{Mohr2011}
\BIBentryALTinterwordspacing
B.~Mohr, \emph{PMPI Tools}.\hskip 1em plus 0.5em minus 0.4em\relax Boston, MA:
  Springer US, 2011, pp. 1570--1575. [Online]. Available:
  \url{https://doi.org/10.1007/978-0-387-09766-4_57}
\BIBentrySTDinterwordspacing

\bibitem{barnes2008regression}
B.~J. Barnes, B.~Rountree, D.~K. Lowenthal, J.~Reeves, B.~De~Supinski, and
  M.~Schulz, ``A regression-based approach to scalability prediction,'' in
  \emph{Proceedings of the 22nd annual international conference on
  Supercomputing}.\hskip 1em plus 0.5em minus 0.4em\relax ACM, 2008, pp.
  368--377.

\bibitem{tang2018vs}
X.~Tang, J.~Zhai, X.~Qian, B.~He, W.~Xue, and W.~Chen, ``vsensor: leveraging
  fixed-workload snippets of programs for performance variance detection,'' in
  \emph{ACM SIGPLAN Notices}, vol.~53, no.~1.\hskip 1em plus 0.5em minus
  0.4em\relax ACM, 2018, pp. 124--136.

\bibitem{terpstra2010collecting}
D.~Terpstra, H.~Jagode, H.~You, and J.~Dongarra, ``Collecting performance data
  with papi-c,'' in \emph{Tools for High Performance Computing 2009}.\hskip 1em
  plus 0.5em minus 0.4em\relax Springer, 2010, pp. 157--173.

\bibitem{hayes2006simulating}
J.~C. Hayes, M.~L. Norman, R.~A. Fiedler, J.~O. Bordner, P.~S. Li, S.~E. Clark,
  M.-M. Mac~Low \emph{et~al.}, ``Simulating radiating and magnetized flows in
  multiple dimensions with zeus-mp,'' \emph{The Astrophysical Journal
  Supplement Series}, vol. 165, no.~1, p. 188, 2006.

\bibitem{rodrigues2011structural}
A.~F. Rodrigues, K.~S. Hemmert, B.~W. Barrett, C.~Kersey, R.~Oldfield,
  M.~Weston, R.~Risen, J.~Cook, P.~Rosenfeld, E.~Cooper-Balis \emph{et~al.},
  ``The structural simulation toolkit,'' \emph{ACM SIGMETRICS Performance
  Evaluation Review}, vol.~38, no.~4, pp. 37--42, 2011.

\bibitem{fischer2008nek5000}
P.~F. Fischer, J.~W. Lottes, and S.~G. Kerkemeier, ``nek5000 web page,'' 2008.

\bibitem{mohr2014scalable}
B.~Mohr, ``Scalable parallel performance measurement and analysis
  tools-state-of-the-art and future challenges,'' \emph{Supercomputing
  frontiers and innovations}, vol.~1, no.~2, pp. 108--123, 2014.

\bibitem{knobloch2020tools}
M.~Knobloch and B.~Mohr, ``Tools for gpu computing--debugging and performance
  analysis of heterogenous hpc applications,'' \emph{Supercomputing Frontiers
  and Innovations}, vol.~7, no.~1, pp. 91--111, 2020.

\bibitem{an2011score}
D.~an~Mey, S.~Biersdorf, C.~Bischof, K.~Diethelm, D.~Eschweiler, M.~Gerndt,
  A.~Kn{\"u}pfer, D.~Lorenz, A.~Malony, W.~E. Nagel \emph{et~al.}, ``Score-p: A
  unified performance measurement system for petascale applications,'' in
  \emph{Competence in High Performance Computing 2010}.\hskip 1em plus 0.5em
  minus 0.4em\relax Springer, 2011, pp. 85--97.

\bibitem{scorephp}
\BIBentryALTinterwordspacing
``Score-p homepage. score-p consortium.'' [Online]. Available:
  \url{http://www.score-p.org}
\BIBentrySTDinterwordspacing

\bibitem{shende2006tau}
S.~S. Shende and A.~D. Malony, ``The tau parallel performance system,''
  \emph{The International Journal of High Performance Computing Applications},
  vol.~20, no.~2, pp. 287--311, 2006.

\bibitem{tauhp}
\BIBentryALTinterwordspacing
``Tau homepage. university of oregon.'' [Online]. Available:
  \url{http://tau.uoregon.edu}
\BIBentrySTDinterwordspacing

\bibitem{muller2007developing}
M.~S. M{\"u}ller, A.~Kn{\"u}pfer, M.~Jurenz, M.~Lieber, H.~Brunst, H.~Mix, and
  W.~E. Nagel, ``Developing scalable applications with vampir, vampirserver and
  vampirtrace.'' in \emph{PARCO}, vol.~15.\hskip 1em plus 0.5em minus
  0.4em\relax Citeseer, 2007, pp. 637--644.

\bibitem{knupfer2008vampir}
A.~Kn{\"u}pfer, H.~Brunst, J.~Doleschal, M.~Jurenz, M.~Lieber, H.~Mickler,
  M.~S. M{\"u}ller, and W.~E. Nagel, ``The vampir performance analysis
  tool-set,'' in \emph{Tools for high performance computing}.\hskip 1em plus
  0.5em minus 0.4em\relax Springer, 2008, pp. 139--155.

\bibitem{vampirhg}
\BIBentryALTinterwordspacing
``Vampir homepage. technical university dresden.'' [Online]. Available:
  \url{http://www.vampir.eu}
\BIBentrySTDinterwordspacing

\bibitem{scalascahg}
\BIBentryALTinterwordspacing
``Scalasca homepage. julich supercomputing centre and german research school
  for simulation sciences.'' [Online]. Available: \url{http://www.scalasca.org}
\BIBentrySTDinterwordspacing

\bibitem{labarta1996dip}
J.~Labarta, S.~Girona, V.~Pillet, T.~Cortes, and L.~Gregoris, ``Dip: A parallel
  program development environment,'' in \emph{European Conference on Parallel
  Processing}.\hskip 1em plus 0.5em minus 0.4em\relax Springer, 1996, pp.
  665--674.

\bibitem{servat2009detailed}
H.~Servat, G.~Llort, J.~Gim{\'e}nez, and J.~Labarta, ``Detailed performance
  analysis using coarse grain sampling,'' in \emph{European Conference on
  Parallel Processing}.\hskip 1em plus 0.5em minus 0.4em\relax Springer, 2009,
  pp. 185--198.

\bibitem{paraver}
\BIBentryALTinterwordspacing
``Paraver homepage. barcelona supercomputing center.'' [Online]. Available:
  \url{http://www.bsc.es/paraver}
\BIBentrySTDinterwordspacing

\bibitem{becker2007automatic}
D.~Becker, F.~Wolf, W.~Frings, M.~Geimer, B.~J. Wylie, and B.~Mohr, ``Automatic
  trace-based performance analysis of metacomputing applications,'' in
  \emph{2007 IEEE International Parallel and Distributed Processing
  Symposium}.\hskip 1em plus 0.5em minus 0.4em\relax IEEE, 2007, pp. 1--10.

\bibitem{zhai2014cypress}
J.~Zhai, J.~Hu, X.~Tang, X.~Ma, and W.~Chen, ``Cypress: combining static and
  dynamic analysis for top-down communication trace compression,'' in
  \emph{SC'14: Proceedings of the International Conference for High Performance
  Computing, Networking, Storage and Analysis}.\hskip 1em plus 0.5em minus
  0.4em\relax IEEE, 2014, pp. 143--153.

\bibitem{noeth2007scalable}
M.~Noeth, F.~Mueller, M.~Schulz, and B.~R. De~Supinski, ``Scalable compression
  and replay of communication traces in massively p arallel e nvironments,'' in
  \emph{2007 IEEE International Parallel and Distributed Processing
  Symposium}.\hskip 1em plus 0.5em minus 0.4em\relax IEEE, 2007, pp. 1--11.

\bibitem{krishnamoorthy2010scalable}
S.~Krishnamoorthy and K.~Agarwal, ``Scalable communication trace compression,''
  in \emph{Proceedings of the 2010 10th IEEE/ACM International Conference on
  Cluster, Cloud and Grid Computing}.\hskip 1em plus 0.5em minus 0.4em\relax
  IEEE Computer Society, 2010, pp. 408--417.

\bibitem{knupfer2005construction}
A.~Knupfer and W.~E. Nagel, ``Construction and compression of complete call
  graphs for post-mortem program trace analysis,'' in \emph{2005 International
  Conference on Parallel Processing}.\hskip 1em plus 0.5em minus 0.4em\relax
  IEEE, 2005, pp. 165--172.

\bibitem{arnold2007stack}
D.~C. Arnold, D.~H. Ahn, B.~R. De~Supinski, G.~L. Lee, B.~P. Miller, and
  M.~Schulz, ``Stack trace analysis for large scale debugging,'' in \emph{2007
  IEEE International Parallel and Distributed Processing Symposium}.\hskip 1em
  plus 0.5em minus 0.4em\relax IEEE, 2007, pp. 1--10.

\bibitem{january2015allinea}
C.~January, J.~Byrd, X.~Or{\'o}, and M.~O’Connor, ``Allinea map: Adding
  energy and openmp profiling without increasing overhead,'' in \emph{Tools for
  High Performance Computing 2014}.\hskip 1em plus 0.5em minus 0.4em\relax
  Springer, 2015, pp. 25--35.

\bibitem{kaufmann2003craypat}
S.~Kaufmann and B.~Homer, ``Craypat-cray x1 performance analysis tool,''
  \emph{Cray User Group (May 2003)}, 2003.

\bibitem{wang2018spindle}
H.~Wang, J.~Zhai, X.~Tang, B.~Yu, X.~Ma, and W.~Chen, ``Spindle: informed
  memory access monitoring,'' in \emph{2018 Annual Technical Conference}, 2018,
  pp. 561--574.

\bibitem{weber2016structural}
M.~Weber, R.~Brendel, T.~Hilbrich, K.~Mohror, M.~Schulz, and H.~Brunst,
  ``Structural clustering: a new approach to support performance analysis at
  scale,'' in \emph{2016 IEEE International Parallel and Distributed Processing
  Symposium}.\hskip 1em plus 0.5em minus 0.4em\relax IEEE, 2016, pp. 484--493.

\bibitem{laguna2015debugging}
I.~Laguna, D.~H. Ahn, B.~R. de~Supinski, T.~Gamblin, G.~L. Lee, M.~Schulz,
  S.~Bagchi, M.~Kulkarni, B.~Zhou, Z.~Chen \emph{et~al.}, ``Debugging
  high-performance computing applications at massive scales,''
  \emph{Communications of the ACM}, vol.~58, no.~9, pp. 72--81, 2015.

\bibitem{zhou2011vrisha}
B.~Zhou, M.~Kulkarni, and S.~Bagchi, ``Vrisha: using scaling properties of
  parallel programs for bug detection and localization,'' in \emph{Proceedings
  of the 20th international symposium on High performance distributed
  computing}.\hskip 1em plus 0.5em minus 0.4em\relax ACM, 2011, pp. 85--96.

\bibitem{laguna2011large}
I.~Laguna, T.~Gamblin, B.~R. de~Supinski, S.~Bagchi, G.~Bronevetsky, D.~H. Anh,
  M.~Schulz, and B.~Rountree, ``Large scale debugging of parallel tasks with
  automaded,'' in \emph{Proceedings of 2011 International Conference for High
  Performance Computing, Networking, Storage and Analysis}.\hskip 1em plus
  0.5em minus 0.4em\relax ACM, 2011, p.~50.

\bibitem{mitra2014accurate}
S.~Mitra, I.~Laguna, D.~H. Ahn, S.~Bagchi, M.~Schulz, and T.~Gamblin,
  ``Accurate application progress analysis for large-scale parallel
  debugging,'' in \emph{ACM SIGPLAN Notices}, vol.~49, no.~6.\hskip 1em plus
  0.5em minus 0.4em\relax ACM, 2014, pp. 193--203.

\bibitem{coarfa2007scalability}
C.~Coarfa, J.~Mellor-Crummey, N.~Froyd, and Y.~Dotsenko, ``Scalability analysis
  of spmd codes using expectations,'' in \emph{Proceedings of the 21st annual
  international conference on Supercomputing}.\hskip 1em plus 0.5em minus
  0.4em\relax ACM, 2007, pp. 13--22.

\bibitem{bohme2010identifying}
D.~Bohme, M.~Geimer, F.~Wolf, and L.~Arnold, ``Identifying the root causes of
  wait states in large-scale parallel applications,'' in \emph{2010 39th
  International Conference on Parallel Processing}.\hskip 1em plus 0.5em minus
  0.4em\relax IEEE, 2010, pp. 90--100.

\bibitem{chen2015critical}
J.~Chen and R.~M. Clapp, ``Critical-path candidates: Scalable performance
  modeling for mpi workloads,'' in \emph{2015 IEEE International Symposium on
  Performance Analysis of Systems and Software}.\hskip 1em plus 0.5em minus
  0.4em\relax IEEE, 2015, pp. 1--10.

\bibitem{yu2020colab}
T.~Yu, P.~Petoumenos, V.~Janjic, H.~Leather, and J.~Thomson, ``Colab: a
  collaborative multi-factor scheduler for asymmetric multicore processors,''
  in \emph{Proceedings of the 18th ACM/IEEE International Symposium on Code
  Generation and Optimization}, 2020, pp. 268--279.

\end{thebibliography}

\end{document}